\documentclass[titlepage,pallis,twoside]{wpallis}
\usepackage{fancyh}
\usepackage[l]{floatflt}
\usepackage{epsfig}
\usepackage{amssymb}
\usepackage{latexsym}
\usepackage{times}
\usepackage{slashed,cite}
\usepackage{upgreek}
\numberwithin{equation}{section}

\newcommand{\ec}{\end{center}}
\newcommand{\bec}{\begin{center}}

\newcommand{\eem}{\end{matrix}}
\newcommand{\bem}{\begin{matrix}}
\newcommand{\eeq}{\end{equation}}
\newcommand{\beq}{\begin{equation}}
\newcommand{\ba}{\begin{array}}
\newcommand{\ea}{\end{array}}
\newcommand{\bea}{\begin{eqnarray}}
\newcommand{\eea}{\end{eqnarray}}
\newcommand{\baq}{\begin{eqnarray}}
\newcommand{\eaq}{\end{eqnarray}}

\newcommand\eqs[2]{Eqs.~(\ref{#1}) and (\ref{#2})}

\newcommand\eqss[3]{Eqs.~(\ref{#1}), (\ref{#2}) and (\ref{#3})}

\newcommand{\ftn}{\footnotesize}

\newcommand{\TeV}{{\mbox{\rm TeV}}}

\newcommand{\GeV}{{\mbox{\rm GeV}}}

\newcommand{\sFref}[2]{Fig.~\ref{#1}-{\ftn\sf ({#2})}}
\newcommand{\sEref}[2]{Eq.~(\ref{#1}{\ftn\sf {#2}})}

\newcommand{\etal}{{\it et al.\/}}

\def\to{\rightarrow}

\def\lf{\left(}
\def\rg{\right)}
\newcommand\vev[1]{\langle {#1} \rangle}
\newcommand{\Gr}{\ensuremath{\widetilde{G}}}
\newcommand{\Yb}{\ensuremath{Y_{B}}}
\newcommand{\Yg}{\ensuremath{Y_{\Gr}}}

\newcommand{\Vjhi}{\ensuremath{V_{\widetilde N\rm I}}}
\newcommand{\Vhi}{\ensuremath{\widehat V_{\widetilde N\rm I}}}
\newcommand{\Hhi}{\ensuremath{\widehat H_{\widetilde N\rm I}}}
\newcommand{\Ohi}{\ensuremath{\Omega_{\widetilde N\rm I}}}
\newcommand{\Khi}{\ensuremath{K_{\widetilde N\rm I}}}
\newcommand{\Shi}{\ensuremath{S_{\widetilde N\rm I}}}
\newcommand{\Vhio}{\ensuremath{\widehat V_{\widetilde N\rm I0}}}

\newcommand{\mP}{\ensuremath{m_{\rm P}}}
\newcommand{\Mpq}{\ensuremath{M_{\rm PQ}}}
\newcommand{\la}{\ensuremath{\lambda_a}}
\newcommand{\lm}{\ensuremath{\lambda_\mu}}

\newcommand{\ck}{\ensuremath{c_\mathcal{R}}}
\newcommand{\kx}{\ensuremath{k_X}}
\newcommand{\kn}{\ensuremath{k_{\widetilde N}}}

\newcommand{\Tpq}{\ensuremath{T_{\rm PQ}}}
\newcommand{\mpq}{\ensuremath{m_{\rm PQ}}}
\newcommand{\Gpq}{\ensuremath{\Gamma_{\rm PQ}}}
\newcommand{\Tsn}{\ensuremath{T_{\sn}}}
\newcommand{\Gsn}{\ensuremath{\Gamma_{\sn}}}
\newcommand{\msn}{\ensuremath{m_{\sn}}}

\newcommand{\hef}{\ensuremath{h_{\rm eff}}}
\newcommand{\deff}{\ensuremath{\delta_{\rm eff}}}

\newcommand{\hd}{{\ensuremath{H_d}}}
\newcommand{\hu}{{\ensuremath{H_u}}}

\newcommand{\ns}{\ensuremath{n_{\rm s}}}
\newcommand{\as}{\ensuremath{\alpha_{\rm s}}}

\newcommand{\rcc}{\ensuremath{\mathcal{R}}}
\newcommand{\rce}{\ensuremath{\widehat{\mathcal{R}}}}
\newcommand{\Ve}{\ensuremath{\widehat{V}}}
\newcommand{\He}{\ensuremath{\widehat{H}}}
\newcommand{\Ne}{\ensuremath{\widehat{N}}}
\newcommand{\sni}{\ensuremath{N^c_i}}
\newcommand{\sn}{\ensuremath{\widetilde N}}

\def\ve{\varepsilon}
\def\aal{{\bar\alpha}}
\def\bbet{{\bar\beta}}
\def\al{{\alpha}}

\def\th{{\theta}}

\newcommand{\Trh}{\ensuremath{T_{\rm rh}}}
\newcommand{\sg}{\ensuremath{\sigma}}

\newcommand{\sgf}{\ensuremath{\sigma_{\rm f}}}
\newcommand{\xsg}{\ensuremath{x_{\sigma}}}

\newcommand{\ld}{\ensuremath{\lambda}}
\newcommand{\Ld}{\ensuremath{\Lambda}}

\newcommand{\se}{\ensuremath{\widehat\sigma}}

\newcommand{\sex}{\ensuremath{\widehat{\sigma}_*}}

\newcommand{\geu}{\ensuremath{\widehat g}}

\newcommand{\vtau}{\ensuremath{\uptau}}

\newcommand{\eL}{{\ensuremath{\varepsilon_{L}}}}

\def\Ka{K\"{a}hler potential}

\def\FHI{non-M$\sn$I~}

\begin{document}

\thispagestyle{empty}
%%%%%%%%%%%%%%%

\title[]{\Large\bfseries\scshape Non-Minimal Sneutrino Inflation,
Peccei-Quinn Phase Transition and non-Thermal Leptogenesis}

\author{\large\bfseries\scshape C. Pallis and N. Toumbas}
\address[] {\sl Department of Physics, University of Cyprus, \\
P.O. Box 20537, CY-1678 Nicosia, CYPRUS}

\begin{abstract}{{\bfseries\scshape Abstract} \\
\par We consider a phenomenological extension of the minimal supersymmetric
standard model which incorporates non-minimal chaotic inflation,
driven by a quartic potential associated with the lightest
right-handed sneutrino. Inflation is followed by a Peccei-Quinn
phase transition based on renormalizable superpotential terms,
which resolves the strong CP and $\mu$ problems of the minimal
supersymmetric standard model provided that one related parameter
of the superpotential is somewhat small. Baryogenesis occurs via
non-thermal leptogenesis, which is realized by the inflaton
decay. Confronting our scenario with the current observational
data on the inflationary observables, the baryon assymetry of the
universe, the gravitino limit on the reheating temperature and the
upper bound on the light neutrino masses, we constrain the effective
Yukawa coupling involved in the decay of the inflaton to
relatively small values and the inflaton mass to values lower than
$10^{12}~\GeV$. }
\hspace*{-1.43cm} \publishedin{{\sl  J. Cosmol.  Astropart. Phys.
} {\bf 02}, 019 (2011)}
\end{abstract}\maketitle

\setcounter{page}{1} \pagestyle{fancyplain}

%\addtolength{\headheight}{.5cm}

\rhead[\fancyplain{}{ \bf \thepage}]{\fancyplain{}{\sl
Non-M$\sn$I, PQPT \& non-Thermal Leptogenesis}}
\lhead[\fancyplain{}{\sl \leftmark}]{\fancyplain{}{\bf \thepage}}
\cfoot{}

\tableofcontents\vskip-1.3cm\noindent\rule\textwidth{.4pt}\\
%\vspace*{.3cm}

\section{Introduction}\label{intro}

Recently \emph{non-minimal inflation} (non-MI) \cite{wmap3}, i.e. inflation
arising in the presence of a non-minimal coupling between the
inflaton field and the Ricci scalar curvature, $\rcc$, has gained
a fair amount of attention \cite{sm1, love, unitarizing,
nmi, ld, linde1, linde2}. In particular, it is shown that
non-minimal chaotic inflation based on a quartic potential
\cite{nmchaotic} with a quadratic non-minimal coupling to gravity
can be realized in both a non-supersymmetric \cite{sm1, love, ld}
and a \emph{sypersymmetric} (SUSY) framework \cite{linde1,
linde2}, provided that the inflaton couples strongly enough to $\rcc$.
In the latter case, the recently developed \cite{linde2}
superconformal approach to \emph{supergravity} (SUGRA) greatly
facilitates the relevant model building. In most of the models
proposed, the inflaton is identified with the Higgs field(s) of
the \emph{Standard Model} (SM) or the next-to-MSSM (\emph{Minimal
SUSY SM}) \cite{linde1,linde2} -- see also \cref{SusyHiggs}.

Motivated by the various attractive features of the MSSM \cite{eg}
-- such as the resolution of the hierarchy problem, the
achievement of gauge coupling unification and the candidature of
the lightest SUSY particle as cold dark matter -- we consider
it as the starting point of our investigation. Despite its
successes, however, the MSSM fails to address a number of
important issues. For instance, the strong CP and $\mu$ problems,
the generation of the observed \emph{baryon asymmetry of the
universe} (BAU) and the existence of tiny but non-zero neutrino
masses are some fundamental issues which remain open within the MSSM.
For the resolution of these, it seems imperative to
supplement the MSSM with additional superfields, which in the simplest
cases are singlets under the SM gauge group, ${G_{\rm SM}}=
SU(3)_{\rm c}\times SU(2)_{\rm L}\times U(1)_{Y}$, so that gauge
coupling unification is not disrupted. Consequently, new
candidates (besides the Higgs boson) for driving non-MI arise.

%Prepei n Sec.2 va 3ekivaei stnv idia selida

In \cref{suzuki, tony, rsym} a resolution to the aforementioned
problems of the MSSM was proposed within a framework that implements a
\emph{Peccei-Quinn symmetry breaking phase transition} (PQPT). In
those models non-renormalizable superpotential terms are added,
involving some singlets that develop \emph{vacuum expectation
values} (VEVs) of the order of the PQ symmetry breaking scale. As
a consequence, the $\mu$ and the strong CP problems \cite{pq} of the MSSM
can be simultaneously solved, and in addition a new intermediate scale arises
which generates Majorana masses for three \emph{right-handed}
(RH) neutrinos, $\sni$. The inclusion of $\sni$ is necessary so
that the smallness of neutrino masses is explained through the
well-known see-saw mechanism \cite{seesaw}. These same superfields
can play an important role in the generation of the BAU via
non-thermal leptogenesis \cite{inlept, chaotic1, baryo}. This
latter attractive possibility is invalidated, though, in the cases
studied in \cref{suzuki, tony}, where the PQPT follows a period of
thermal inflation \cite{thermalI} that leads to a very low
reheating temperature. An enormous entropy production occurs,
diluting any preexisting, non-thermally created, lepton asymmetry.
This dilution can be avoided if we adopt the scheme of \cref{rsym}
but then, the PQ field cannot be zero during inflation -- see
below.

On the other hand, non-thermal leptogenesis can be enhanced if the
scalar component, $\sn$, of the lightest $\sni$ is the inflaton
itself as firstly proposed in \cref{murayama}. In this case the
branching ratio of the inflaton decay (which now triggers
leptogenesis) into a lepton plus a Higgs boson is \cite{baryo}
maximized. However, $\sn$-inflation, in its simplest realization
\cite{tony, murayama, sneutrino1}, is of the chaotic type -- for
other scenarios see \cref{sneutrinoF, sneutrinoD} -- and
therefore, trans-Planckian inflaton-field values are typically
required to allow for a sufficiently long period of inflation. The
implementation of inflation then necessitates the adoption of
special types of \Ka, as in \cref{chaotic1}, so that SUGRA
corrections are kept under control -- for other proposals related
to chaotic inflation with a quadratic potential, see
\cref{sneutrino2}. Moreover, minimal chaotic inflation driven by a
quartic potential seems \cite{circ} to be ruled out by the fitting
to the seven-year data of the \emph{Wilkinson Microwave Anisotropy
Probe Satellite} (WMAP7), \emph{baryon-acoustic-oscillations}
(BAO) and \emph{Hubble constant} ($H_0$) data \cite{wmap}.

In this paper we construct a model of \emph{non-minimal $\sn$
inflation} (\FHI) retaining the successful ingredients of the
picture above. To this aim, $\sn$ (the lightest RH sneutrino) is
coupled to one of the PQ fields, which can be confined to zero
during inflation -- see \cref{linde2}. We then show that the model
naturally leads to non-MI within SUGRA, provided that a particular
parameter of the superpotential is sufficiently small.
Sub-Planckian values of the inflaton field are allowed in a wide
range of the parameter space, and the adopted type of \Ka\ is more
or less well-motivated. Also the inflationary observables turn out
to lie within the current data. The \FHI is followed by a PQPT
driven by renormalizable superpotential terms as in \cref{goto,
pqhi}, whereas the $\mu$ parameter of the MSSM can be generated
from the PQ scale via a non-renormalizable term as in
\cref{suzuki, tony, rsym}. The reheating temperature is determined
exclusively by the decay of $\sn$ and is high enough ($>100~\GeV$)
so that non-perturbative electroweak sphalerons are operative and,
consequently, non-thermal \cite{inlept} leptogenesis and the
subsequent generation of the BAU can be realized. As usually in
similar models -- cf. \cref{baryo, Ndomination, sneutrino1,
sneutrino2, sneutrinoF, sneutrinoD} -- consistency with the
constraint on the gravitino ($\Gr$) abundance \cite{gravitino,
brand, kohri} requires a relatively small effective Yukawa
coupling constant $\lf10^{-8}-10^{-3}\rg$. The smallness of this
coupling though may be explained through a broken flavor symmetry
\cite{baryo, Ndomination}.

Below, we present the basic ingredients of our model
(Sec.~\ref{fhim}) and describe the inflationary (Sec.~\ref{fhi})
and post-inflationary dynamics (Sec.~\ref{pfhi}). We then restrict
the parameters of our model (Sec.~\ref{cont}) and summarize our
conclusions (Sec.~\ref{con}). Details concerning the formulation
of non-minimally coupled scalar fields within SUGRA are presented
in the Appendix. Throughout the text, we use natural units for
Planck's constant, Boltzmann's constant and the speed of light
($\hbar=c=k_{\rm B}=1$); the subscript of type $,\chi$ denotes
derivation \emph{with respect to} (w.r.t.) the field $\chi$ (e.g.,
$_{,\chi\chi}=\partial^2/\partial\chi^2$); charge conjugation is
denoted by a star and $\log~[\ln]$ stands for logarithm with basis
$10~[e]$. Finally, we follow the conventions of \cref{kolb} for
the quantities related to the gravitational sector of our model.

\section{Model Description}\label{fhim}

We focus on a PQ invariant extension of the MSSM, inspired by
\cref{suzuki, tony}, which links the generation of intermediate
masses for RH neutrinos, $\sni$, with a PQPT. Besides the (color)
anomalous PQ symmetry $U(1)_{\rm PQ}$, the model possesses also an
anomalous $R$ symmetry $U(1)_{R}$, and the baryon number symmetry
$U(1)_B$. The PQ symmetry $U(1)_{\rm PQ}$ is spontaneously broken
at the PQ breaking scale $f_a\sim\lf10^{10}-10^{12}\rg~{\rm GeV}$
(which coincides with the axion decay constant -- for a review see
Ref.~\cite{kim}) via the VEVs acquired by two $G_{\rm SM}$ singlet
left-handed superfields $\bar X$ and $X$. The representations
under $G_{\rm SM}$ and the charges under the global symmetries of
the various matter and Higgs superfields are listed in
Table~\ref{table}.

\renewcommand{\arraystretch}{1.2}

\begin{table}[!t]
\begin{center}
\begin{tabular}{|c|c|c|c|c|}\hline
{\sc Super-}&{\sc Representations}&\multicolumn{3}{|c|}{\sc Global
Charges}
\\\cline{3-5}{\sc fields}&{\sc under $G_{\rm SM}$}& \hspace*{0.2cm}$R$\hspace*{0.2cm} &
\hspace*{0.2cm}{PQ}\hspace*{0.2cm} &{$B$}
\\\hline
\multicolumn{5}{|c|}{\sc Matter Fields}\\\hline
{$L_i$} &{$({\bf 1, 2}, -1/2)$}& $0$ & $-3$ &$0$
\\
{$e^c_i$} & {$({\bf 1, 1}, 1)$} &$2$&{$1$}&{$0$}
\\
{$\sni$} & {$({\bf 1, 1}, 0)$} &$2$&{$1$}&{$0$}
\\
{$Q_i$} &{$({\bf 3, 2}, 1/6)$}& $1$ & $-1$ &$1/3$
 \\
{$u^c_i$} & {$({\bf \bar 3, 1}, -2/3)$} &$1$ &{$-1$}&$-1/3$
\\
{$d^c_i$} & {$({\bf \bar 3, 1}, 1/3)$} &$1$ &{$-1$}&$-1/3$
\\ \hline
\multicolumn{5}{|c|}{\sc Higgs Fields}
\\\hline
{$\hd$} & {$({\bf 1, 2}, -1/2)$}&$2$ &$2$ &$0$ \\
{$\hu$} & {$({\bf 1, 2}, 1/2)$}&$2$ &$2$ &$0$ \\ \hline
{$P$} &{$({\bf 1, 1}, 0)$}&{$4$}&{$0$} & {$0$}\\
{$\bar X$}&$({\bf 1, 1}, 0)$&{$0$}&{$2$}&{$0$}\\
{$X$}&$({\bf 1, 1}, 0)$&{$0$}&{$-2$}&{$0$}
\\ \hline
\end{tabular}
\end{center}
\vchcaption[]{\sl \small The representations under $G_{\rm SM}$
and the extra global charges of the superfields of our
model.}\label{table}
\end{table}

In particular, the superpotential, $W$, of our model naturally
splits into two parts:
\beq W=W_{\rm MSSM}+W_{\rm NPQ}, \eeq
where $W_{\rm MSSM}$ is the part of $W$ which contains the usual
terms -- except for the $\mu$ term -- of the MSSM, supplemented by
Yukawa interactions among the left-handed leptons and $\sni$:
\beq W_{\rm MSSM} = h_{Eij} {e}^c_i {L}_j \hd+h_{Dij} {d}^c_i
{Q}_j \hd + h_{Uij} {u}^c_i {Q}_j \hu+ h_{Nij} \sni L_j \hu.
\label{wmssm}\eeq
Here, the group indices have been suppressed; the $i$-th generation
$SU(2)_{\rm L}$ doublet left-handed quark and lepton superfields
are denoted by $Q_i$ and $L_i$ respectively, whereas the
$SU(2)_{\rm L}$ singlet antiquark [antilepton] superfields by
$u^c_i$ and ${d_i}^c$ [$e^c_i$ and $\sni$] respectively. The
electroweak Higgs superfields, which couple to the up [down] quark
superfields, are denoted by $\hu$ [$\hd$].

On the other hand, $W_{\rm NPQ}$ is the part of $W$ which is
relevant for non-M\sn I, the generation of the Majorana masses for
$\sni$, the spontaneous breaking of ${\rm U(1)}_{\rm PQ}$ and the
generation of the $\mu$ term of the MSSM. It takes the form
\beq\label{Whi} W_{\rm NPQ}= \ld_{i} X\sni \sni+\la P(\bar
XX-f^2_a/4)+\lm{X^2\hu\hd\over \mP}, \eeq
where $m_{\rm P}\simeq 2.44\cdot 10^{18}~{\rm GeV}$ is the reduced
Planck scale and $P$ is a $G_{\rm SM}$ singlet left-handed
superfield involved in the breaking of $U(1)_{\rm PQ}$.
The parameters $\la$ and $f_a$ are made positive by field
redefinitions. Moreover, we chose a basis in the $N_i-N_j$ space
where the coupling constant matrix $\lambda$ is real and diagonal.
In order to produce the CP-violation necessary for leptogenesis,
we include three $\sni$. Assuming that $\sni$ are strongly
hierarchical, i.e. $\ld_1=\ld\ll\ld_2,\ld_3$, the scalar
components of the two heavier $\sni$ roll to their minima fairly
quickly, since their potential is steeper - especially if we
further assume that these are minimally coupled to gravity in
contrast to the lightest one. Thus the scalar component, $\sn$, of
the lightest of the $\sni$'s controls the relevant slow-roll
dynamics and it can therefore be identified as the inflaton. On
the other hand, the outcome of leptogenesis is governed by the one
from $\sni$ with the smallest decay rate. In the following, we
concentrate on the case where $\sn$ drives both \FHI and
leptogenesis. Therefore, the three generation model can be
simplified to an effective one-generation model in the $\sni$
sector, with the only remnant of the other two generations being a
non-vanishing CP-asymmetry for leptogenesis -- see \cref{murayama,
sneutrinoF}. Henceforth we suppress family indices.

According to our general discussion in the Appendix -- see
\Eref{Kg} -- the implementation of \FHI within SUGRA requires the
adoption of a frame function, $\Ohi$, related to the
K\"ahler potential, $\Khi$, of the following form
\beq
\Ohi=-3e^{-\Khi/3\mP^2}=-3+{|\sn|^2\over\mP^2}+{|P|^2\over\mP^2}+{|\bar
X|^2\over\mP^2}+{|X|^2\over\mP^2}-{\kx}{|X|^4\over\mP^4}-{3\kn\over4\mP^2}\lf
\sn^2+\sn^{*2}\rg,\label{minK}\eeq
where the complex scalar components of the superfields $P, \bar X$
and $X$ are denoted by the same symbol and the coefficients $\kx$
and $\kn$ are taken, for simplicity, real. Comparing this
expression with \Eref{Omg}, we remark that we adopt the standard
non-minimal coupling for the inflaton, $\sn$, i.e. ${\rm F}=\kn
\sn^2/4\mP^2$, and we added the sixth term in the \emph{RH side}
(RHS) in order to cure the tachyonic mass problem encountered in
similar models \cite{linde1, linde2} -- see \Sref{fhi1}. Note that
F breaks explicitly the imposed $R$ and PQ symmetries and causes a
dependence of $\Ohi$ on the phase $\th$ of $\sn$ which can be
written as $\sn=\sg e^{i\th}/\sqrt{2}$. As we show in \Sref{fhi1},
the model admits stable inflationary trajectories along which
$\th$ is stabilized at $\th=0$. When $\th\sim0$, the choice
$\kn>1$ ensures the positivity of the scale function, $-\Ohi/3$,
even for relatively large values of $\sg$.

In the limit where $\mP$ tends to infinity, the matter sector
decouples from gravity, and we can obtain the SUSY limit, $V_{\rm
SUSY}$, of the SUGRA potential, $\Vhi$. This turns out to be
\beq\label{VF} V_{\rm
SUSY}=\left(4\ld^2|\sn|^2+\la^2\left|P\right|^2\right)|X|^2
+\left|\ld\sn^2+\la P\bar X\right|^2+ \la^2\left|\bar X
X-f^2_a/4\right|^2.\eeq
From the potential in Eq.~(\ref{VF}), we find that the SUSY vacuum
lies at
\beq\vev{\sn}=0,\>\>\>
\vev{P}\simeq0\>\>\>\mbox{and}\>\>\>|\langle\phi_{X}\rangle|=2|\vev{X}|=2|\vev{\bar
X}|=f_a,\label{vevs} \eeq
where we have introduced the canonically normalized scalar field
$\phi_{X}=2X=2\bar X$. Note that, since the sum of the arguments
of $\vev{\bar X}$, $\vev{X}$ must be $0$, $\bar X$ and $X$ can be
brought to the real axis by an appropriate PQ transformation.
Moreover, after including soft SUSY breaking terms, $\vev{P}$ can
become \cite{goto} of order $1~{\rm TeV}$ and the minimization of
$V_{\rm F}$ in Eq.~(\ref{VF}) requires that $\vev{\bar
X}=\vev{X}$. Needless to say that for field values greater than
$f_a$, the $\TeV$-scale soft SUSY breaking terms can be safely
ignored during the cosmological evolution. After the spontaneous
breaking of $U_{\rm PQ}(1)$, the first term of the \emph{RH side}
(RHS) of \Eref{Whi} generates intermediate scale masses for the
$\sni$'s $M_i\sim\ld_if_a$ and, thus, seesaw masses \cite{seesaw}
for the light neutrinos, whereas the third of the RHS of
\Eref{Whi} leads to the $\mu$ term of the MSSM, with
$|\mu|\sim\lambda_\mu\left|\vev{X}\right|^2/m_{\rm P}$. Both
scales are of the right magnitude if
$\left|\vev{X}\right|=f_a/2\simeq 5\cdot 10^{11}~{\rm GeV}$,
$\ld_1\sim 1$ and $\lambda_\mu\sim(0.001-0.01)$.

In conclusion, $W_{\rm NPQ}$ leads to a spontaneous breaking of
$U_{\rm PQ}(1)$. The same superpotential $W_{\rm NPQ}$ also gives
rise to a stage of \FHI and a PQPT, as analyzed in
\Sref{fhi}. An indication for such a possibility can be seen
by examining $V_{\rm SUSY}$ in \Eref{VF}, which becomes
\beq V_{\rm SUSY}=\ld^2\sn^4+\la^2 f_a^2/4\>\>\>\mbox{along the
direction}\>\>\>X=\bar X=0.\label{flat}\eeq
Clearly, for $\sn\gg f_a$, $V_{\rm SUSY}$ tends to a quartic
potential. Therefore, $W_{\rm NPQ}$ can be employed in conjunction
with $\Khi$ in \Eref{minK} for the realization of \FHI along the
lines of \cref{linde2}. Moreover, for lower $\sn$'s, $V_{\rm
SUSY}$ takes an almost constant value, which can drive a PQPT.

It should be mentioned that the non-minimal gravitational
coupling, instanton and soft SUSY breaking effects explicitly
break $U(1)_R\times U(1)_{\rm PQ}$ to a discrete subgroup. It is
then important to ensure that this subgroup is not spontaneously
broken by $\vev{X}$ and $\vev{\bar X}$, since otherwise
cosmologically disastrous domain walls are produced \cite{sikivie}
during the PQPT. Note that $U(1)_{R}\times U(1)_{\rm PQ}$ is also
broken during \FHI due to the non-zero $\sn$, but it is restored in
the SUSY vacuum. The explicitly unbroken subgroup of
$U(1)_{R}\times U(1)_{\rm PQ}$ can be deduced from the solutions of
the system
\beq \label{expb} 4r+2p =0~\lf\mbox{\ftn\sf mod}~2\pi\rg,\>\>\>4r
=0~\lf\mbox{\ftn\sf mod}~2\pi\rg\>\>\>\mbox{and}\>\>\>
-12(r+p)=0~\lf\mbox{\ftn\sf mod}~2\pi\rg, \eeq
where $r$ and $p$ are the phases of a $U(1)_{R}$ and $U(1)_{\rm
PQ}$ rotation respectively. Here we took into account that:
{\ftn\sf (a)} the $R$ [PQ] charge of the langrangian term caused
by the non-minimal gravitational coupling is 4 [2]; {\ftn\sf (b)}
the $R$ charge of $W$ and, thus, of all the soft SUSY breaking
terms, is 4 and {\ftn\sf (c)} the sum of the $R$ [PQ] charges of
the $SU(3)_{\rm c}$ triplets and antitriplets is $-12$ [$-12$]. We
conclude, therefore, that the explicitly unbroken subgroup is
$\mathbb{Z}_4\times \mathbb{Z}_{2}$. It is then easy to check that
this subgroup is not spontaneously broken by $\vev{\bar X}$ and
$\vev{X}$, since the relevant condition
\beq\label{sunb} 2p=0~\lf\mbox{\ftn\sf mod}~2\pi\rg, \eeq
is satisfied automatically as a result of \Eref{expb}. Consequently,
cosmologically disastrous domain walls are not produced during the PQPT
and so, there is no need to further extend \cite{georgi} the
particle content of the model -- cf. \cref{pqhi}.

\section{The Inflationary Epoch}\label{fhi}

\subsection{Structure of the Inflationary
Action}\label{fhi1}

Inserting \Eref{minK} into \Eref{Sfinal}, we can write the action
of our model in the JF (Jordan frame) as follows
\beq \label{Snpq} \Shi=\int d^4x\sqrt{-g}\lf
{1\over6}\mP^2\Ohi\rcc +\delta_{\al\bbet}\partial_\mu\phi^\al
\partial^\mu \phi^{*\bbet}-\Ohi{\cal A}_\mu{\cal
A}^\mu/\mP^2-\Vjhi\rg ,\eeq
with $\phi^\al=\sn, P, X$ and $\bar X$. Also $\Vjhi=\Ohi^2\Vhi/9$
where $\Vhi$ is the EF (Einstein frame) F--term SUGRA scalar
potential, which can be obtained from $W_{\rm NPQ}$ in
Eq.~(\ref{Whi}) -- without the last term of the RHS -- and $\Khi$
in Eq.~(\ref{minK}) by applying \Eref{Vsugra}. Along the direction
$P=X=\bar X=0$, $\Vhi$ and $\Ohi =-3f$ take the forms
\beq \label{Vhi}\Vhio =\mP^4\frac{\ld^2\xsg^4+ 4 \la^2
\Mpq^4}{4f^2}\>\>\>\mbox{with}\>\>\>f=1-{\xsg^2\over6}+\lf{1\over6}+\ck\rg\xsg^2\cos2\th\>\>\>
\mbox{and}\>\>\>\ck=-\frac{1}{6}+\frac{k_{\sn}}{4}\cdot\eeq
Here $\Mpq=f_a/2\mP$ and $\xsg=\sg/\mP$. Recall also that we set
$\sn=\sigma e^{i\th}/\sqrt{2}$. From \Eref{Vhi}, we can easily
verify -- see also the small fluctuations analysis below -- that
for given $\sg$, $\th=0$ (modulo $\pi$) minimizes $\Vhi$. For
$\th=0$ and $\ck\gg1$, $\Shi$ in \Eref{Snpq} takes a form suitable
for the realization of non-M\sn I. Then we can set ${\cal
A}^\mu=0$, and more importantly $\Vhi$ develops a plateau -- note
that $\Mpq\ll1$. The constant potential energy density $\Vhio$ and
the corresponding Hubble parameter $\He_{\rm \sn I0}$ along the
trajectory for which \FHI can take place are given by
\beq \Vhio=
{\ld^2\sg^4\over4f^2}\simeq{\ld^2\mP^4\over4\ck^2}\>\>\>\mbox{and}\>\>\>
\He_{\sn\rm
I0}={\Vhio^{1/2}\over\sqrt{3}\mP}\simeq{\ld\mP\over2\sqrt{3}\ck}
\cdot\label{Vhio}\eeq

In order to check the stability of the direction $P=X=\bar
X=\th=0$ w.r.t. the fluctuations of the fields $\th, P, X$ and
$\bar X$, we expand the latter three in real and imaginary parts
as follows
\bea X= {x_1+ix_2\over\sqrt{2}},\>\>\>\bar X= {\bar x_1+i\bar
x_2\over\sqrt{2}}\>\>\>\mbox{and}\>\>\> P=
{p_1+ip_2\over\sqrt{2}}\cdot \label{cannor} \eea
Performing a Weyl transformation as described in \Eref{weyl}, we
obtain \cite{conformal}
\bea \nonumber S_{\sn\rm I}&=&\int d^4 x
\sqrt{-\geu}\left(-\frac12 \mP^2
\rce+{1\over2}\lf\frac{1}{f}+\frac{3f^2_{,\sg}}{2f^2}\mP^2\rg\geu^{\mu\nu}\lf
\partial_\mu\sg \partial_\nu \sg+\sg^2\partial_\mu\th \partial_\nu
\th\rg\right.
\\ && \left.+\frac{1}{2f}\geu^{\mu\nu}\sum_{\chi}\partial_\mu\chi \partial_\nu
\chi-\Vhi\right), \label{Sni1} \eea
with $\chi=x_1, x_2, \bar x_1, \bar x_2, p_1, p_2$, and we also
take into account that $f_{,\chi}\ll f_{,\sg}$ and $f_{,\th}\ll
f_{,\sg}$ for $\th\sim0$. Note that we keep only terms up to
quadratic order in the fluctuations $\th ,\chi$ and their
derivatives in \Eref{Sni1}. Along the trajectory $P=X=\bar
X=\th=0$, $f$ becomes a function of $\sg$, and so we can introduce
the EF canonically normalized fields, $\se, \widehat \th$ and
$\widehat \chi$, as follows \cite{nmi,linde2}
\beq \label{VJe}
\left(\frac{d\se}{d\sigma}\right)^2=J^2=\frac{1}{f}+{3\over2}\mP^2\left({f_{,\sigma}\over
f}\right)^2,\>\>\>\widehat \th =
J\sg\th\>\>\>\mbox{and}\>\>\>\widehat \chi =\frac{\chi}{\sqrt{f}}
\cdot\eeq
Taking into account the approximate expressions for $\dot\sg$ --
where the dot denotes derivation w.r.t. the cosmic time $t$ -- $J$
and the slow-roll parameters $\widehat\epsilon, \,\widehat\eta$,
which are displayed in \Sref{fhi2}, we can verify that, during a
stage of slow-roll non-M\sn I, $\dot{\widehat \th}\simeq J\sg\dot
\th$ since $J\sg\simeq\sqrt{6}\mP$, and $\dot{\widehat
\chi}\simeq\dot \chi/\sqrt{f}$. For the latter, the quantity $\dot
f/f^{3/2}$, involved in relating $\dot{\widehat \chi}$ to $\dot
\chi$, turns out to be negligibly small, since $\dot f/
f^{3/2}=f_{,\sg}\dot\sg/
f^{3/2}=-{\ld\sqrt{\widehat\epsilon|\widehat\eta|}\mP/2\sqrt{3}\ck}$.
Therefore the action in \Eref{Sni1} takes the form
\beq S_{\sn\rm I}=\int d^4 x \sqrt{-\geu}\left(-\frac12 \mP^2
\rce+\frac12\geu^{\mu\nu} \sum_{\phi}\partial_\mu\widehat\phi
\partial_\nu \widehat\phi-\Vhi\right), \label{Sni} \eeq
where $\phi$ stands for $\sg, \th, x_1, x_2, \bar x_1, \bar x_2,
p_1$ and $p_2$.

Along the inflationary path, we can easily check that the first
derivatives of $\Vhi$ w.r.t. $\phi$ are equal to zero. The
curvature of $\Vhi$ w.r.t. $\widehat\th$ can be studied separately
since $\partial^2\Vhi/\partial\widehat\th\partial\widehat\chi=0$.
Moreover the stability of the path $P=X=\bar X=\th=0$ w.r.t. the
fluctuations of $\th$ is automatic, since the mass squared of
$\widehat\th$, $m_{\widehat\th}^2$, turns out to be positive.
Indeed we find
\beq \label{mth}
m_{\widehat\th}^2={\ld^2\mP^2\lf1+6\ck\rg\xsg^4\over
6J^2f^3}\simeq{\ld^2\mP^2\over 3\ck^3}=4\He_{\sn\rm I0}^2.\eeq
The two $3\times 3$ mass squared matrices
$M_{A}^2=\lf\partial^2\Vhi/\partial\chi_\al\partial\chi_\beta\rg$,
with $A=1$ and $\chi_\al=\widehat p_1, \widehat x_1, \widehat{\bar
x}_1$ or $A=2$ and $\chi_\al=\widehat p_2, \widehat x_2,
\widehat{\bar x}_2$, have the following eigenvalues
\beq \label{ms}  m_{\widehat x}^2=\ld^2\mP^2\xsg^2{ \lf 12 +
\xsg^2 f_1\rg\lf6 k_X f-1\rg\over 6f^2f_1} \>\>\>\mbox{and}\>\>\>
m_{\widehat
y_\pm}^2=\ld\mP^2\xsg^2{\lf\ld\pm3\la\ck\rg\xsg^2\pm3\la\over 6
f^2} \eeq
corresponding to eigenstates $\widehat x_1$ (or $\widehat x_2$)
and $\widehat y_{1\pm}=\lf\widehat p_1\pm\widehat{\bar
x}_1\rg/\sqrt{2}~\lf\mbox{or}~\widehat y_{2\pm}=\lf\widehat
p_2\pm\widehat{\bar x}_2\rg/\sqrt{2}\rg$ respectively. The
considered inflationary trajectory, $P=X=\bar X=\th=0$, is a
stable valley of local minima, provided that $m_{\widehat
x}^2\geq0$ and $m_{\widehat y_\pm}^2\geq0$, i.e.
\beq \label{scr}  \mbox{\sf\ftn (a)}\>\>\>\sg\gtrsim \sg_{\rm
1c}={\mP\over\sqrt{6\kx\ck}} \>\>\>\mbox{and}\>\>\>\mbox{\sf\ftn
(b)}\>\>\>\sg\geq \sg_{\rm 2c}=\mP\sqrt{{3\la\over\ld
-3\la\ck}}\>\>\>\mbox{with}\>\>\>\la<\ld/3\ck. \eeq
In practice the condition of \sEref{scr}{b} is much more
restrictive than \sEref{scr}{a} for $\kx\sim1$. Indeed, from
\Eref{ms}, it is evident that $\kx\gtrsim1$ assists us to achieve
$m_{\widehat x}^2>0$ -- in accordance with the results of
\cref{linde2}. On the other hand, given that for $\sg<\mP$, we
need $\ck\gg1$, \sEref{scr}{b} requires a clear hierarchy between
$\ld$ and $\la$, e.g. for $\ck\simeq10^2$, we need
$\la/\ld\lesssim10^{-3}$. This ratio can be slightly increased
(almost one order of magnitude) if we include additional terms
such as $k_{X \tilde N}|\sn|^2|X|^2$ or $k_{P \tilde
N}|\sn|^2|P|^2$ in the \Ka\ in \Eref{minK}. Since the resulting
increase of $\la$ has no significant impact on our results, we
choose to stick with the most minimal possible \Ka\ needed for the
viability of our model, avoiding more complications. We have also
numerically verified that $m_{\widehat\th}\geq\Hhi$, $m_{\widehat
x}\geq\Hhi$ and $m_{\widehat y_\pm}\geq\Hhi$, during the last
$50-60$ e-foldings of non-M\sn I, and so any inflationary
perturbations of the fields $\widehat\th,\>\widehat x_{1,2}$ and
$\widehat y_{1,2\pm}$ are safely eliminated.

The constant tree-level potential energy density in \Eref{Vhio}
causes SUSY breaking, leading to the generation of one-loop
radiative corrections, which can be calculated by employing the
well-known Coleman-Weinberg formula \cite{cw}. We find
\beq V_{\rm rc}={1\over 64\pi^2}\lf
m_{\widehat\th}^4\ln{m_{\widehat\th}^2\over\Lambda^2}+2m_{\widehat
x}^4\ln{m_{\widehat x}^2\over\Lambda^2}+ 2m_{\widehat
y_+}^4\ln{m_{\widehat y_+}^2\over\Lambda^2}+2m_{\widehat
y_-}^4\ln{m_{\widehat y_-}^2\over\Lambda^2}-4\widetilde
m^4\ln{\widetilde m^2\over\Lambda^2}\rg\eeq
where $\Lambda$ is a renormalization mass scale and $\widetilde
m={\sqrt{2}\lambda\mP\xsg/f^{3/2}}$ is the eigenvalue of the
fermion matrices. As we verified numerically, $V_{\rm rc}$ has no
significant effect on the inflationary dynamics. This is because
the slope of the inflationary path is generated at the classical
level -- see the expressions for $\widehat\epsilon$ and
$\widehat\eta$ below -- and so, the contribution of $V_{\rm rc}$
to $\Vhi$ remains subdominant.

Based on the action of \Eref{Sni} with $\Vhi\simeq\Vhio+V_{\rm
rc}$, we can proceed to the analysis of \FHI in the EF, using the
standard slow-roll approximation \cite{review, lectures}. It can
be shown \cite{general} that the results calculated this way are
the same as if we had calculated them using the non-minimally
coupled scalar field in the JF.

\subsection{The Inflationary Observables}\label{fhi2}

According to our analysis above, when \Eref{scr} is satisfied, the
universe undergoes a period of slow-roll \FHI, which is determined
by the condition -- see e.g. \cref{review, lectures}:
\numparts\baq \nonumber && \>\>\>\>\>\>\>\>\>\>\>\>\>\>\>\>\>\>
{\ftn\sf max}\{\widehat\epsilon(\sigma),|\widehat\eta(\sigma)|\}\leq1,\>\>\>\mbox{where}\\
&&  \label{sr1}\widehat\epsilon=
{\mP^2\over2}\left(\frac{\Ve_{{\rm \sn I},\se}}{\Ve_{\rm \sn
I}}\right)^2={\mP^2\over2J^2}\left(\frac{\Ve_{{\rm \sn
I},\sigma}}{\Ve_{\rm \sn I}}\right)^2\simeq {4\mP^4\over3\ck^2\sg^4}\\
\mbox{and}\>\>\> && \label{sr2}\widehat\eta= m^2_{\rm
P}~\frac{\Ve_{{\rm \sn I},\se\se}}{\Ve_{\rm \sn I}}={\mP^2\over
J^2}\left( \frac{\Ve_{{\rm \sn I},\sigma\sigma}}{\Ve_{\rm \sn
I}}-\frac{\Ve_{{\rm \sn I},\sigma}}{\Ve_{\rm \sn I}}{J_{,\sg}\over
J}\right)\simeq-{4\mP^2\over3\ck\sg^2}\cdot \eaq\endnumparts
\hspace*{-.145cm}
Here we employ \Eref{Vhio} and the following approximate
relations:
\beq J\simeq \sqrt{6}\mP/\sg,\>\>\>\widehat V_{\sn\rm
I,\sg}\simeq\ld^2\mP^6/\ck^3\sg^3\>\>\>\mbox{and}\>\>\>\widehat
V_{\sn\rm I,\sg\sg}\simeq-3\ld^2\mP^6/\ck^3\sg^4.\eeq
The numerical computation reveals that \FHI terminates due to the
violation of the $\widehat\epsilon$ criterion at a value of $\sg$
equal to $\sgf$, which is calculated to be
\beq \widehat\epsilon\lf\sgf\rg=1\>\>\>\Rightarrow\>\>\>
\sgf=\lf{4/3}\rg^{1/4}{\mP/\sqrt{\ck}}\cdot \label{sgap}\eeq
Note, in passing, that for $\sg\geq\sgf$ the evolution of
$\widehat\sg$ -- or $\sg$ via \Eref{VJe} -- is governed by the
equation of motion
\beq 3\Hhi\dot {\hat{\sg}}=-\widehat V_{\sn \rm
I,\hat{\sg}}\>\Rightarrow\>3\Hhi J^2\dot {\sg}=-\widehat V_{\sn\rm
I,\sg}\>\Rightarrow\>\dot\sg=-{\ld\mP^3/3\sqrt{3}\ck^2\sg}.
\label{eqf}\eeq
Using \eqss{sr1}{sr2}{eqf}, we can derive the expression for $\dot
f/f^{3/2}$ given above \Eref{Sni}.

The number of e-foldings, $\widehat N_{*}$, that the scale
$k_{*}=0.002/{\rm Mpc}$ suffers during non-M\sn I can be
calculated through the relation
\begin{equation}
\label{Nhi}  \widehat{N}_{*}=\:\frac{1}{m^2_{\rm P}}\;
\int_{\se_{\rm f}}^{\se_{*}}\, d\se\: \frac{\Ve_{\rm \sn
I}}{\Ve_{\rm\sn I,\se}}= {1\over\mP^2}\int_{\sigma_{\rm
f}}^{\sigma_{*}}\, d\sigma\: J^2\frac{\Ve_{\rm \sn I}}{\Ve_{\rm
\sn I,\sigma}},
\end{equation}
where $\sigma_*~[\se_*]$ is the value of $\sg~[\se]$ when $k_*$
crosses the inflationary horizon. Given that
$\sgf\ll\sg_*$, we can write $\sg_*$ as a function of
$\widehat{N}_{*}$ as follows
\beq \label{s*}
\widehat{N}_{*}\simeq{3\ck\over4\mP^2}\lf\sg_*^2-\sgf^2\rg\>\>\Rightarrow\>\>\sg_*=2\mP\lf{\widehat
N_*\over 3\ck}\rg^{1/2}\cdot\eeq

The power spectrum $P_{\cal R}$ of the curvature perturbations
generated by $\sigma$ at the pivot scale $k_*$ is estimated as
follows
\begin{equation}  \label{Prob}
P^{1/2}_{\cal R}=\: \frac{1}{2\sqrt{3}\, \pi\mP^3} \;
\frac{\Ve_{\rm \sn I}(\sex)^{3/2}}{|\Ve_{\rm \sn
I,\se}(\sex)|}=\frac{|J(\sigma_*)|}{2\sqrt{3}\, \pi\mP^3} \;
\frac{\Ve_{\rm \sn I}(\sigma_*)^{3/2}}{|\Ve_{\rm \sn
I,\sigma}(\sigma_*)|}\simeq{\ld\sg_*^2\over8\sqrt{2}\pi\mP^2}\simeq
{\sqrt{2}\ld\widehat{N}_{*}\over12\pi\ck}, \eeq
where \Eref{s*} is employed to derive the last equality of the relation
above. At the same pivot scale, we can also calulate the (scalar)
spectral index, $n_{\rm s}$, its running, $a_{\rm s}$, and the
scalar-to-tensor ratio, $r$, via the relations:
\numparts\baq \label{ns} && n_{\rm s}=\: 1-6\widehat\epsilon_*\ +\
2\widehat\eta_*\simeq1-{2/\widehat N_*}, \>\>\> \\
&& \label{as} \alpha_{\rm s}
=\:{2\over3}\left(4\widehat\eta_*^2-(n_{\rm
s}-1)^2\right)-2\widehat\xi_*\simeq{-2/\widehat N^2_*}\>\>\> \\
\mbox{and}\>\>\> && \label{rt}
r=16\widehat\epsilon_*\simeq{12/\widehat N^2_*}, \eaq\endnumparts
\hspace*{-.2cm}
where $\widehat\xi=\mP^4 {\Ve_{\rm \sn I,\widehat\sigma} \Ve_{\rm
\sn I,\widehat\sigma\widehat\sigma\widehat\sigma}/\Ve_{\rm \sn
I}^2}= \mP\,\sqrt{2\widehat\epsilon}\,\widehat\eta_{,\sigma}/
J+2\widehat\eta\widehat\epsilon$ and the variables with subscript
$*$ are evaluated at $\sigma=\sigma_{*}$. Comparing the results of
this section with the observationally favored values, we constrain
the parameters of our model in \Sref{cont}.

\section{The Post-Inflationary Evolution}\label{pfhi}

A complete SUSY inflationary scenario should specify the
transition to the radiation dominated era and also explain the
origin of the observed BAU consistently with the $\Gr$ constraint.
These goals can be accomplished within our set-up, as we describe
in this section. The basic features of the post-inflationary era
of our model are exhibited in \Sref{lept}. A more precise analysis
of the evolution during this era can be obtained by solving
numerically the relevant Boltzmann equations, as in \Sref{Boltz}.
Finally, useful analytical expressions reproducing accurately our
results are presented in \Sref{Anal}.

\subsection{The General Set-up}\label{lept}

When \FHI is over, $\sn$ undergoes a very short period of
fast-roll until it reaches its critical value $\sg_{2\rm
c}/\sqrt{2}$. Afterwards, $\sn$ and the PQ system (comprised by
$X, \bar X$ and $P$) fall to their SUSY minimum values, acquiring
masses $\msn$ and $\mpq$ respectively, which can be computed from
$V_{\rm SUSY}$ in \Eref{VF}. These are given by
\beq \label{masses} \mbox{\sf\ftn (a)}\>\>\> \msn=\ld
f_a\>\>\>\mbox{and}\>\>\>\mbox{\sf\ftn (b)}\>\>\>\mpq=\la
f_a/\sqrt{2}.\eeq
Note that due to hierarchy between $\ld$ and $\la$ established in
\sEref{scr}{b}, $\msn>\mpq$. Consequently, the post-inflationary
energy density of the universe is dominated by the $\sn$
condensate which undergoes a phase of damped oscillations about
the SUSY vacuum, when $H\simeq\msn$, and decays \cite{murayama}
predominantly into $\widetilde\hu + L$ or $H_u^* + \widetilde
L^*$, via the tree-level couplings derived from the last term in
the RHS of Eq.~(\ref{wmssm}). The initial energy density of this
oscillatory system is estimated by $\rho_{\rm
1i}\simeq3\mP^2\msn^2$, corresponding to $H_{\rm 1i}\simeq\msn$.
The decay temperature of $\sn$, $\Tsn$, which coincides with the
reheating temperature, $\Trh$, in our model -- see below -- is
\cite{quin} given by
\beq \label{T1rh} \Tsn= c_T\sqrt{\Gsn
\mP}\>\>\>\mbox{with}\>\>\>c_T=\left(72\over5\pi^2g_{*}\right)^{1/4},\eeq
where $g_{*}$ counts the effective number of relativistic degrees
of freedom at temperature $T_{\sn}$. We find $g_{*}\simeq240$ for
the MSSM spectrum plus the particle content of the superfields
$P$, $\bar X$ and $X$. Also $\Gsn$ is the decay width of $\sn$
given by
\beq\Gsn={1\over4\pi}\hef
^2\,\msn\>\>\>\mbox{where$\>\>\>\hef=\sqrt{\sum_i
\left|h_{N1i}\right|^2}$} \label{gammas}\eeq
is an effective Yukawa coupling, linked to the light
neutrino masses, and can be considered as a free parameter.

The aforementioned two channels for the $\sn$ decay have different
branching ratios when CP conservation is violated. Interference
between tree-level and one-loop diagrams generates a lepton-number
asymmetry \cite{kolb, baryo} which, for a normal hierarchical mass
spectrum of light neutrinos, reads
\beq\label{el} \ve_L = \frac {3}{8\pi}\frac{m_{\nu 3}}{\langle
\hu\rangle ^2}\msn \deff\,. \eeq
Here $|\deff|\leq1$, which is treated as a free parameter in our
approach, represents the magnitude of CP violation; $m_{\nu 3}$ is
the heaviest neutrino mass and we take $\vev{\hu}=174~\GeV$
(adopting the large $\tan\beta$ regime). If $\Trh<\msn$, the
out-of-equilibrium condition \cite{baryo} for the implementation
of leptogenesis is automatically satisfied. The resulting
lepton-number asymmetry after reheating can be partially converted
through sphaleron effects into baryon-number asymmetry. However,
the required $\Trh$ must be compatible with constraints for the
$\Gr$ abundance, $Y_{\Gr}$, at the onset of nucleosynthesis.

On the other hand, the system consisting of the two complex scalar
fields $P$ and $(\delta\bar X+\delta X)/\sqrt{2}$ (where
$\delta\bar X=\bar X-f_a/2$ and $\delta X=X-f_a/2$) enters into an
oscillatory phase about the PQ minimum and eventually decays, via
the non-renormalizable coupling in the RHS of Eq.~(\ref{Whi}), to
Higgses and Higgsinos with a common decay width $\Gpq$
\cite{lectures} and a corresponding decay temperature $T_{\rm PQ}$
given by
\beq T_{\rm PQ}=
c_T\sqrt{\Gpq\mP}\>\>\>\mbox{where}\>\>\>\Gpq={1\over2\pi}\lm^2\left({f_a\over
2\mP}\right)^2m_{\rm PQ}. \label{T2rh}\eeq
Note that due to the hierarchy between $\ld$ and $\la$ in
\Eref{scr}, the decay of $X$ to $\sni$'s is kinematically
forbidden. Due to the same fact, $\Tpq$ turns out to be quite
suppressed, and so a possible domination of the PQ oscillatory
system could dilute any preexisting $Y_L$ and $\Yg$. However,
$\vev{X}\ll \mP$ -- in contrast to the VEVs of moduli occurring in
superstring theory \cite{thermalI} which are of the order of $\mP$
-- and therefore, the initial energy density of the $X$ and $\bar
X$'s oscillations, $\rho_{\rm 2i}$, is reduced w.r.t. the energy
density of the universe at the onset of these oscillations,
$\rho_{H\rm PQ}$ -- cf. \cref{baryo, Ndomination}. Namely
\beq \rho_{\rm 2i}\simeq\mpq^2 |\vev{X}|^2\ll \rho_{H\rm PQ} =
3\mP^2\mpq^2\>\>\>\mbox{for}\>\>\> H \simeq \mpq.\label{r2i}\eeq
This is a crucial point since it assists us to avoid any dilution
of the produced $Y_L$ (and $Y_{\Gr}$) at $T=T_{\sn}$, as we show
in the following.

\subsection{The Relevant Boltzmann Equations}\label{Boltz}

The energy density, $\rho_1$ [$\rho_2$], of the oscillatory system
with decay width $\Gsn$ [$\Gpq$], the energy density of produced
radiation, $\rho_{\rm R}$, the number density of lepton asymmetry,
$n_L$, and the one of $\Gr$, $n_{\Gr}$, satisfy the following
Boltzmann equations -- cf.~\cref{pqhi, gpp, kohri, kolb,
murayama}:
\numparts\baq%
&& \dot \rho_1+3H\rho_1+\Gsn \rho_1=0,\label{nf}\\
&& \dot\rho_2+3H\rho_2+\Gpq\rho_2=0,\label{nfb} \\
&& \dot\rho_{\rm R}+4H\rho_{\rm R}-
\Gamma_{\sn}\rho_1-\Gpq\rho_2=0,\label{rR}\\
&& \dot n_L+3Hn_L-\ve_L\Gsn\rho_1/\msn=0,\label{nl}\\
&& \dot n_{\Gr}+3Hn_{\Gr}-C_{\Gr} \lf n^{\rm eq}\rg^2=0.\label{ng}
\eaq\endnumparts \hspace{-.14cm}
Here $n^{\rm eq}={\zeta(3)T^3/\pi^2}$ is the equilibrium number
density of the bosonic relativistic species; $C_{\Gr}$ is a
collision term for $\Gr$ production which, in the limit of the
massless gauginos, turns out to be \cite{kohri, brand}
\beq C_{\Gr} = \frac{3\pi}{16\zeta(3)\mP^2}\sum_{i=1}^{3} c_i
g_i^2 \ln\left({k_{i}\over
g_i}\right)\>\>\>\mbox{where}\>\>\>\left\{\bem (c_i)=(33/5,27,72)
\hfill \cr (k_i)=(1.634,1.312,1.271) \hfill \eem \right.\eeq
and $g_i$ (with $i=1,2,3$) are the gauge coupling constants of
the MSSM calculated as functions of the temperature $T$. The latter
quantity and the entropy density, $s$, can be obtained through the
relations
\beq \rho_{\rm R}=\frac{\pi^2}{30}g_*
T^4\>\>\>\mbox{and}\>\>\>s=\frac{2\pi^2}{45}g_* T^3.
\label{rs}\eeq
Also the Hubble expansion parameter, $H$, during this period is
given by
\begin{equation} \label{Hini}
H=\frac{1}{\sqrt{3}\mP} \left(m_{{\Gr}}n_{{\Gr}}+\rho_1
+\rho_2+\rho_{\rm R} \right)^{1/2}.
\end{equation}
Clearly, in the limit of massless MSSM gauginos, the $n_{\Gr}$
computation is $m_{\Gr}$-independent.

The numerical integration of Eqs.~(\ref{nf})--(\ref{ng}) is
facilitated by absorbing the dilution terms. To this end, we find
it convenient to define \cite{quin} the following dimensionless
variables
\begin{equation} \label{fdef}
f_1=\rho_1 R^3,~f_2=\rho_2 R^3,~f_{\rm R}=\rho_{\rm R}
R^4,~f_{L}=n_L R^3\>\>\mbox{and}\>\>\>f_{\Gr}=n_{\Gr} R^3.
\end{equation}
Converting the time derivatives to derivatives w.r.t.
$\vtau=\ln\left(R/R_{\rm i}\right)$, with $R_{\rm i}$ being the
value of the scale factor at the onset of the $\sn$ oscillations
-- the precise value of $R_{\rm i}$ turns out to be numerically
irrelevant -- Eqs.~(\ref{nf})--(\ref{ng}) become
\numparts\baq
Hf^\prime_1&=&-\Gsn f_1,\label{ff}\\
Hf^\prime_2&=&-\Gpq f_2,\label{ffb}\\
Hf^\prime_{\rm R}&=&\Gsn f_1 R+\Gpq f_2 R, \label{fR}\\
Hf^\prime_{L}&=&\eL\Gsn R^3,\label{fL}\\
Hf^\prime_{\Gr}&=&C_{\Gr}\lf n^{\rm eq}\rg^2R^3.\label{fg}
\eaq\endnumparts \hspace{-.14cm}
Also $H$ and $T$ can be expressed in terms of the variables in
Eq.~(\ref{fdef}) as
\beq \label{H2exp}  H=\frac{\sqrt{m_{\Gr}f_{\Gr}+f_1 +f_2 +f_{\rm
R}/R}}{\sqrt{3R^3}\mP}\>\>\>\mbox{and}\>\>\> T=\lf{{30\ f_{\rm
R}\over\pi^2 g_* R^4}}\rg^{1/4}\cdot\eeq
The system of Eqs.~(\ref{ff})--(\ref{fg}) can be solved, imposing
the following initial conditions (the quantities below are
considered functions of the independent variable $\vtau$):
\beq\rho_1(0)=\rho_{1\rm i},\>\>\>\rho_{\rm
R}(0)=n_{\Gr}(0)=n_{L}(0)=0\>\>\>\mbox{and}\>\>\>\rho_2(\vtau_{H\rm
PQ})=\rho_{2\rm i}, \label{init} \eeq
where $\vtau_{H\rm PQ}$ is the value of $\vtau$ corresponding to
the temperature $T_{H\rm PQ}$ which is defined as the solution of
the equation $H\lf T_{H\rm PQ}\rg=\mpq$ and can be found
numerically. Needless to say that we set $\rho_2(\vtau)=0$
for $\vtau<\vtau_{H\rm PQ}$.

\begin{figure}[!t]\vspace*{-.15in}
\hspace*{-.19in}
\begin{minipage}{8in}
\epsfig{file=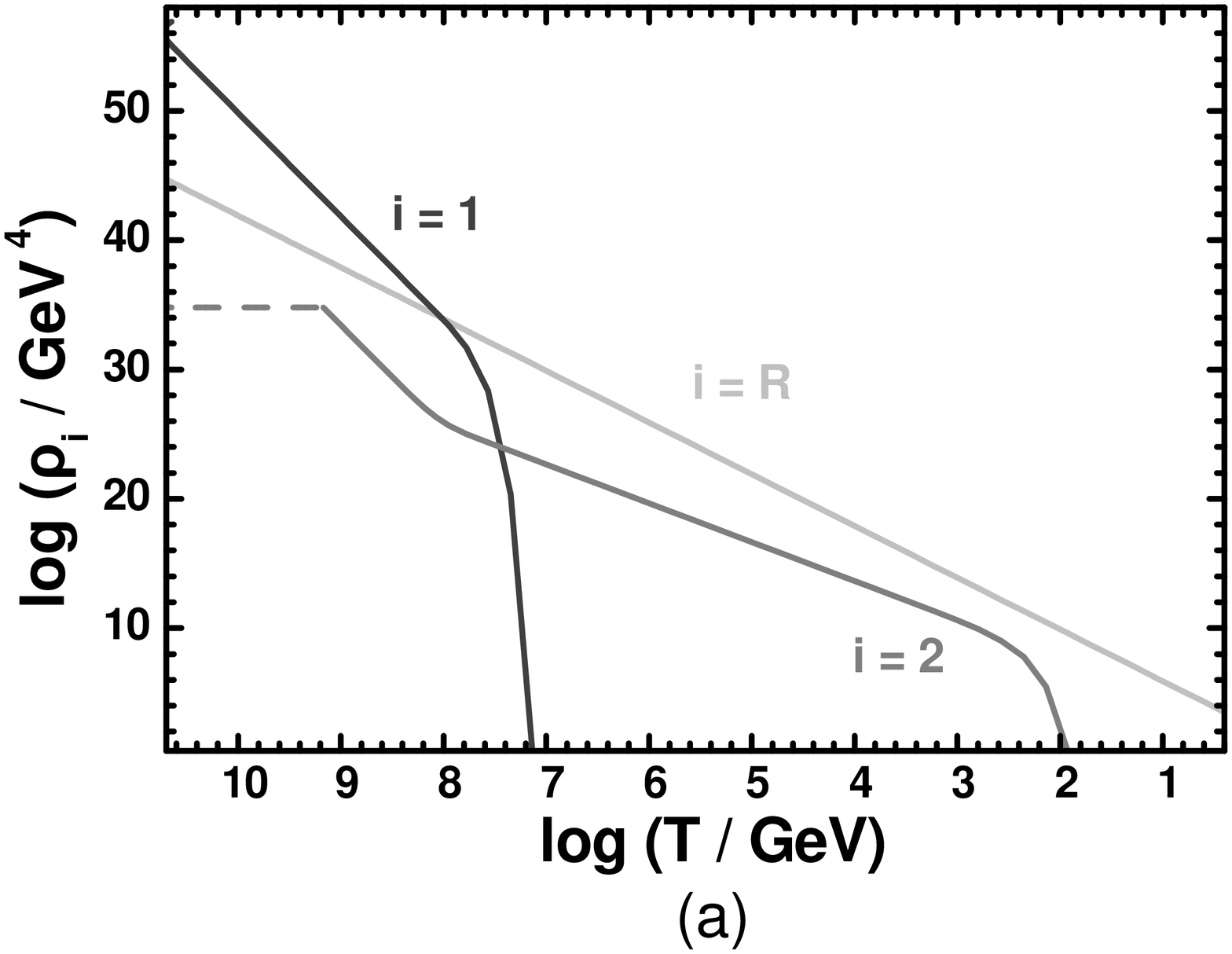,height=3.6in,angle=-90}
\hspace*{-1.2cm}
\epsfig{file=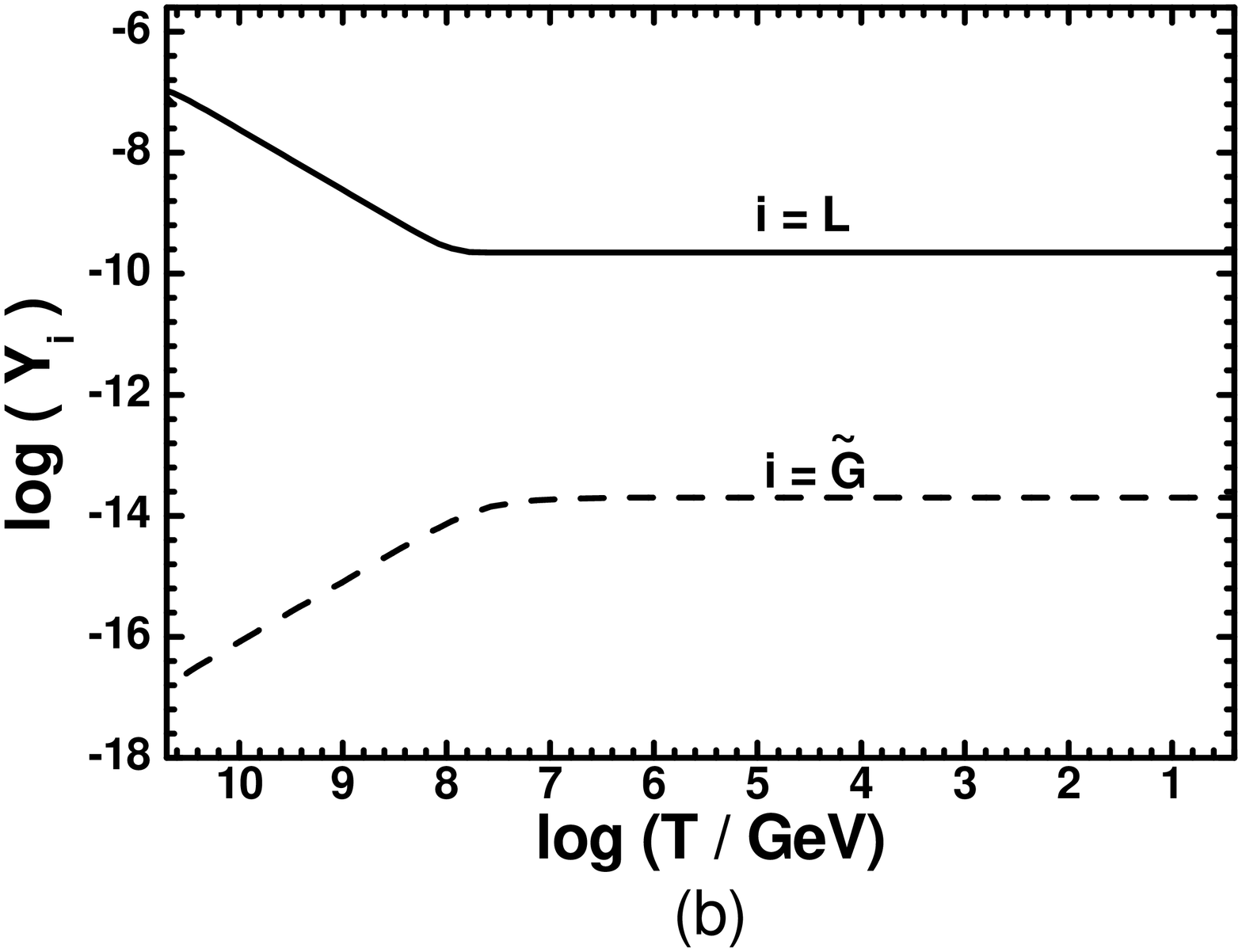,height=3.6in,angle=-90} \hfill
\end{minipage}
\hfill \vchcaption[]{\sl\small \label{fig1} The evolution of the
quantities $\log\rho_i$ with $i=1$ (dark gray line), $i=2$ (gray
line), $i={\rm R}$ (light gray line), $\log\rho_{2\rm i}$ (dashed
gray line), $\log Y_L$ (black solid line) and $\log\Yg$ (black
dashed line) as functions of $\log T$, for $\ld=0.0071,
\la=10^{-6}, f_a=10^{12}~\GeV, ~\mu=1~\TeV,~\hef=10^{-5},~\kx=1$
and $\ck\simeq307$, resulting to $\Yg=2\cdot10^{-14}$ and
$Y_L=2.5\cdot10^{-10}$ for $m_{\nu3}=0.05~{\rm eV}$ and
$\deff=0.01$.}
\end{figure}

In Fig.~\ref{fig1}, we illustrate the cosmological evolution of
the quantities $\log\rho_i$ with $i=1$ (dark gray line), $i=2$
(gray line), $i={\rm R}$ (light gray line), $\log\rho_{2\rm i}$
(dashed gray line), $\log Y_L$ (black solid line) and $\log\Yg$
(black dashed line) as functions of $\log T$ for values of the
parameters which are allowed by all the restrictions described in
\Sref{cont}. In particular we set $\ld=0.0071, \la=10^{-6},
f_a=10^{12}~\GeV,~\mu=1~\TeV,~\hef=10^{-5},~\kx=1,
~\ck\simeq307,~m_{\nu3}=0.05~{\rm eV}$ and $\deff=0.01$. From
Fig.~\ref{fig1}--{\sf\ftn (a)}, we observe that \FHI is followed
by a \emph{matter dominated} (MD) era, due to the oscillating and
decaying inflaton system, which lasts until $T\simeq T_{\sn}$
given by Eq.~(\ref{T1rh}). The completion of the reheating process
corresponds to the intersection of $\rho_1$ with $\rho_{\rm R}$.
Afterwards the universe enters the conventional \emph{radiation
dominated} (RD) epoch of standard Big Bang cosmology. This becomes
possible thanks to \Eref{r2i}, since then the decay of the PQ
system commences at $T=T_{H\rm PQ}$ and is completed at $T\simeq
T_{\rm PQ}$ given by Eq.~(\ref{T2rh}), before its domination over
radiation. So a second episode of reheating does not occur. As we
show below, this is a generic feature of our model. Due to this
fact, the $L$ and ${\Gr}$ yields, $Y_{L}=n_{L}/s$ and
$Y_{\Gr}=n_{{\Gr}}/s$ respectively, take their actual values
($2.5\cdot10^{-10}$ and $2\cdot10^{-14}$) for $T\simeq T_{\sn}$ as
shown in Fig.~\ref{fig2}--{\sf\ftn (b)}. Both numerical values are
compatible with observational data -- see \Sref{cont}.

\subsection{Analytical Approach}\label{Anal}

The numerical findings above can be understood by some simple
analytic formulas. Most of them are widely employed in the
literature -- cf. \cref{sneutrino1, sneutrino2, sneutrinoF,
sneutrinoD, Ndomination, baryo}. In particular, the $B$ yield can
be computed as
\beq {\sf\ftn
(a)}\>\>\>Y_B=-0.35Y_L\>\>\>\mbox{with}\>\>\>{\sf\ftn
(b)}\>\>\>Y_L=n_L/s=c_L T_{\sn} \>\>\>\mbox{where}\>\>\>
c_L=-5\ve_L/4\msn.\label{Yb}\eeq
The numerical factor in \sEref{Yb}{a} is due to sphaleron
effects, whereas the unusual numerical factor in the definition of
$c_L$ is due to the slightly different calculation \cite{quin} of
$\Tsn$ -- cf.~\cref{baryo}. On the other hand, the $\Gr$ yield at
the onset of nucleosynthesis is estimated to be
\beq\label{Ygr} Y_{\Gr}=n_{\Gr}/s\simeq c_{\Gr}
T_{\sn}\>\>\>\mbox{with}\>\>\>c_{\Gr}=1.9\cdot10^{-22}/\GeV.\eeq
Both \eqs{Yb}{Ygr} calculate the correct values of the $B$ and
$\Gr$ abundances provided that no entropy production occurs for
$T<\Tsn$. We show in the following that this is the case for
our model.

The evolution of the various energy densities involved in the
post-inflationary dynamics can be well approximated -- see e.g.
\cref{kolb, quin} -- by the expressions
\numparts\baq \label{r1a} &\rho_1=\rho_{1\rm
i}e^{-3\vtau}&\mbox{for}\>\>\>T\geq\Tsn,\\
 &\rho_2=\rho_{2\rm i}e^{-3\lf\vtau-\vtau_{H\rm
PQ}\rg}&\mbox{for}\>\>\>T\geq\Tpq,\label{r2a} \\
 &\rho_{\rm R}=\rho_{\rm
R}(T_{\sn})\left(T/T_{\sn}\right)^4&\mbox{for}\>\>\>T\leq\Tsn.
\label{rRa}\eaq\endnumparts \hspace*{-.14cm}
Possible domination of $\rho_2$ may occur for $T\leq\Tsn$, since
$\rho_{\rm R}$ is steeper than $\rho_2$. The equality between
these two energy densities could be attained for $T=T_{\rm eq}$
where
\begin{equation} \label{trg} \rho_{\rm R}(T_{\rm eq })=\rho_2(T_{\rm eq
})~\Rightarrow~T_{\rm eq}\simeq T_{\sn}\ {\rho_{2 \rm
i}\over\rho_{H\rm PQ}}={\Mpq^2\over3} T_{\sn}\cdot\end{equation}
In deriving the equation above, we use the fact that $\rho_{\rm
R}(T_{\sn})=\rho_1(T_{\sn})$ given by \Eref{r1a} and that
$e^\vtau\sim T^{-1}$ for $T_{\rm eq}\leq T\leq T_{\sn}$ due to the
isentropic expansion. Also we assume that for $T\geq T_{H\rm PQ}$
we have a MD era driven by $\rho_1$. This is a natural assumption
since the condition of the out-of-equilibrium decay of $\sn$ gives
an upper bound on $\hef$, which prevents the unlimited increase of
$\Tsn$. Indeed
\beq \msn\geq\Tsn\>\Rightarrow\>\hef\leq{2\sqrt{\pi}\over
c_T}\sqrt{\frac{\msn}{\mP}}={2\sqrt{2\ld\pi}\over
c_T}\sqrt{\Mpq},\label{wash}\eeq
where \eqs{T1rh}{masses} are employed. The domination of $\rho_2$
over the several energy densities -- and therefore, a second
reheating process -- can be avoided if we impose the condition
\beq \Tpq\geq T_{\rm
eq}\>\Rightarrow\>\la\geq\ld\hef^2\Mpq^2/3\sqrt{2}\lm^2.\label{Teq1}\eeq
Combining the last relation with \Eref{scr}, we arrive at
$\hef\leq \sqrt{3\sqrt{2}}\lm/\Mpq\ck$. Taking e.g.
$\lm\simeq0.01$, $\ck\simeq10^2$ and $\Mpq\simeq10^{-6}$ the last
relation implies $\hef\lesssim100$ which is much less restrictive
than \Eref{wash} which gives $\hef\lesssim0.01$ for $\ld\leq3.5$
and $c_T\simeq0.3$ -- see \Sref{num}. Therefore, the decay of the
PQ system before its domination over radiation can be naturally
accommodated within our set-up. Moreover, we remark that the
results on $\Tsn,~ Y_L$ and $\Yg$ are independent of $\rho_{H\rm
PQ}$ and $\rho_{\rm 2i}$ or $\mu, \la$ and $f_a$ when \Eref{Teq1}
holds.

\section{Constraining the Model Parameters}\label{cont}

We exhibit the constraints that we impose on our cosmological
set-up in \Sref{cont1}, and delineate the allowed parameter space
of our model in Sec.~\ref{num}.

\subsection{Imposed Constraints}\label{cont1}

Under the assumption that {\sf (i)} the curvature perturbations
generated by $\sigma$ are solely responsible for the observed
curvature perturbations and {\sf (ii)} the violation of \Eref{scr}
occurs after the violation of the slow-roll conditions in
\eqs{sr1}{sr2}, the parameters of our model can be restricted once
we impose the following requirements:

\paragraph{5.1.1} According to the inflationary paradigm,
the horizon and flatness problems of the standard Big Bag
cosmology can be successfully resolved provided that
$\widehat{N}_{*}$ defined by \Eref{Nhi} takes a certain value,
which depends on the details of the cosmological scenario.
Employing standard methods \cite{nmi, hinova}, we can easily
derive the required $\widehat{N}_{*}$ for our model, consistently
with the fact that the PQ oscillatory system remains subdominant
during the post-inflationary era. Namely we obtain
\begin{equation}  \label{Ntot}
\widehat{N}_{*}\simeq22.5+2\ln{V_{\rm\sn
I}(\sg_{*})^{1/4}\over{1~{\rm GeV}}}-{4\over 3}\ln{V_{\rm\sn
I}(\sg_{\rm f})^{1/4}\over{1~{\rm GeV}}}+ {1\over3}\ln {T_{\rm
rh}\over{1~{\rm GeV}}}+{1\over2}\ln{f(\sg_{\rm f})\over
f(\sg_*)}\cdot
\end{equation}

\paragraph{5.1.2} The inflationary observables derived in
\Sref{fhi2} are to be consistent with the fitting \cite{wmap} of
the WMAP7, BAO and $H_0$ data. As usually, we adopt the central
value of $P^{1/2}_{\cal R}$, whereas we allow the remaining
quantities to vary within the 95$\%$ \emph{confidence level}
(c.l.) ranges. Namely,
\begin{equation}  \label{obs}
\mbox{\ftn\sf (a)}\>\>\>P^{1/2}_{\cal R}\simeq4.93\cdot
10^{-5},\>\>\>\mbox{\ftn\sf (b)}\>\>\>n_{\rm
s}=0.968\pm0.024,\>\>\>\mbox{\ftn\sf (c)}\>\>-0.062\leq a_{\rm
s}\leq0.018 \>\>\>\mbox{and}\>\>\>\mbox{\ftn\sf (d)}\>\>r<0.24.
\end{equation}

\paragraph{5.1.3} For the realization of \FHI, we assume that $\ck$ takes relatively
large values -- see e.g. \Eref{Sni1}. This assumption may
\cite{cutoff, unitarizing} jeopardize the validity of the
classical approximation, on which the analysis of the inflationary
behavior is based. To avoid this inconsistency -- which is rather
questionable \cite{cutoff, linde2} though -- we have to check the
hierarchy between the ultraviolet cut-off, $\Ld=\mP/\ck$, of the
effective theory and the inflationary scale, which is represented
by $\Vhi(\sg_*)^{1/4}$ or, less restrictively, by the
corresponding Hubble parameter, $\widehat
H_*=\Vhi(\sg_*)^{1/2}/\sqrt{3}\mP$. In particular, the validity of
the effective theory implies \cite{cutoff}
\beq \label{Vl}\mbox{\ftn\sf (a)}\>\>\> \Vhi(\sg_*)^{1/4}\leq\Ld
\>\>\>\mbox{or}\>\>\>\mbox{\ftn\sf (b)}\>\>\> \widehat
H_*\leq\Ld\>\>\>\mbox{for}\>\>\>\mbox{\ftn\sf
(c)}\>\>\>\ck\geq1.\eeq

\paragraph{5.1.4} In agreement with our assumption about hierarchical light
neutrino masses and the results of neutrino oscillation
experiments \cite{neutrino}, $m_{\nu 3}$ -- involved in the
definition of $\ve_L$ in \Eref{el} -- can be related to the
squared mass difference measured in atmospheric neutrino
oscillations, $\Delta m^2_\oplus$. Taking the central value of the
latter quantity, we set
\beq m_{\nu 3}\simeq\sqrt{\Delta
m^2_\oplus}=\lf2.43\cdot10^{-3}\rg^{1/2}~{\rm eV}\simeq0.05~{\rm
eV}. \label{matm}\eeq
This value is low enough to ensure that the lepton asymmetry is
not erased by lepton number violating $2\to2$ scatterings
\cite{erasure} at all temperatures between $\Tsn$ and $100~\GeV$.

\paragraph{5.1.5} The interpretation of BAU through non-thermal
leptogenesis dictates \cite{wmap} at 95\% c.l.
\beq Y_B=\lf8.74\pm0.42\rg\cdot10^{-11}\>\Rightarrow\>8.32\leq
Y_B/10^{-11}\leq9.16.\label{BAUwmap}\eeq
Given our ignorance about $\deff$ in \Eref{el}, we impose only the
lower bound of the inequality above as an absolute constraint.

\paragraph{5.1.6} In order to avoid spoiling the success of the
SBB nucleosynthesis, an upper bound on $Y_{\Gr}$ is to be imposed
depending on the $\Gr$ mass, $m_{\Gr}$, and the dominant $\Gr$
decay mode. For the conservative case where $\Gr$ decays with a
tiny hadronic branching ratio, we have \cite{kohri}
\beq  \label{Ygw} Y_{\Gr}\lesssim\left\{\bem
%\begin{array}{rl}
10^{-15}\hfill \cr
10^{-14}\hfill \cr
10^{-13}\hfill \cr
10^{-12}\hfill \cr\eem
%\end{array}
\right.\>\>\>\mbox{for}\>\>\>m_{\Gr}\simeq\left\{\bem
%\begin{array}{rl}
0.45~{\rm TeV}\hfill \cr
0.69~{\rm TeV}\hfill \cr
10.6~{\rm TeV}\hfill \cr
13.5~{\rm TeV.}\hfill \cr\eem
%\end{array}
\right.\eeq
The bound above can be somehow relaxed in the case of a stable
$\Gr$. However, it is achievable in our model, as we see below.

\subsection{Results}\label{num}

As can be easily seen from the relevant expressions above, our
cosmological set-up depends on the following parameters:
$$\ld,\>\la,\>f_a,\>\lm,\>\kx,\>\ck,\>\hef\>\>\>\mbox{and}\>\>\>\deff.$$
Our results are independent of $\la$ and $\kx$, provided that
\eqs{scr}{Teq1} are satisfied. With these conditions, the
contribution of $V_{\rm rc}$ to $\Vhi$ remains subdominant and the
PQ system decays before radiation domination. We therefore set
$\la=10^{-6}$ and $\kx=1$ throughout our calculation. The chosen
$\la$ is close to its largest value allowed by \sEref{scr}{b},
whereas $\kx$ is fixed to a natural value. Also $\deff$ affects
exclusively the $Y_L$ caclulation through \eqs{el}{Yb}. We take
$\deff=1$, which allows us to obtain via \Eref{el} the maximal
\cite{Ibara} possible lepton asymmetry. This choice in conjuction
with the imposition of the lower bound on $\Yb$ in \Eref{BAUwmap}
provides the most conservative restriction on our parameters.
Also, we set $\lm=0.01$ [$\lm=1$] so as to obtain $\mu\sim1~\TeV$
with $f_a=10^{12}~\GeV$ [$f_a=10^{11}~\GeV$] -- evidently, the
generation of the $\mu$ term of the MSSM through the PQ symmetry
breaking does not favor lower values for $f_a$. As we show below,
the selected values for the above quantities give us a wide and
natural allowed region for the remaining fundamental parameters
($\ld,~\ck$ and $\hef$) of our model. In our numerical code, we
use as input parameters $\sigma_*, \hef, f_a$ and $\ck$. For every
chosen $\ck\geq1$ and $\hef$, we restrict $\ld$ and $\sigma_*$ so
as the conditions \Eref{Ntot} -- with $\Trh$ evaluated
consistently using \Eref{T1rh} -- and (\ref{obs}{\ftn\sf a}) are
satisfied. Let us also remark that in our numerical calculations,
we use the complete formulas for the slow-roll parameters and
$P_{\cal R}^{1/2}$ in \eqss{sr1}{sr2}{Prob} and not the
approximate relations, which are listed in \Sref{fhi2} for the
sake of presentation.

%%%%%%%%%%%%%%%%%%%%%%%%%%%%%%%%%%%%%%%%%%%%%%%%%%%%%%%%%%%%%%%%%%%%%
\begin{figure}[!t]\vspace*{-.15in}
\hspace*{-.19in}
\begin{minipage}{8in}
\epsfig{file=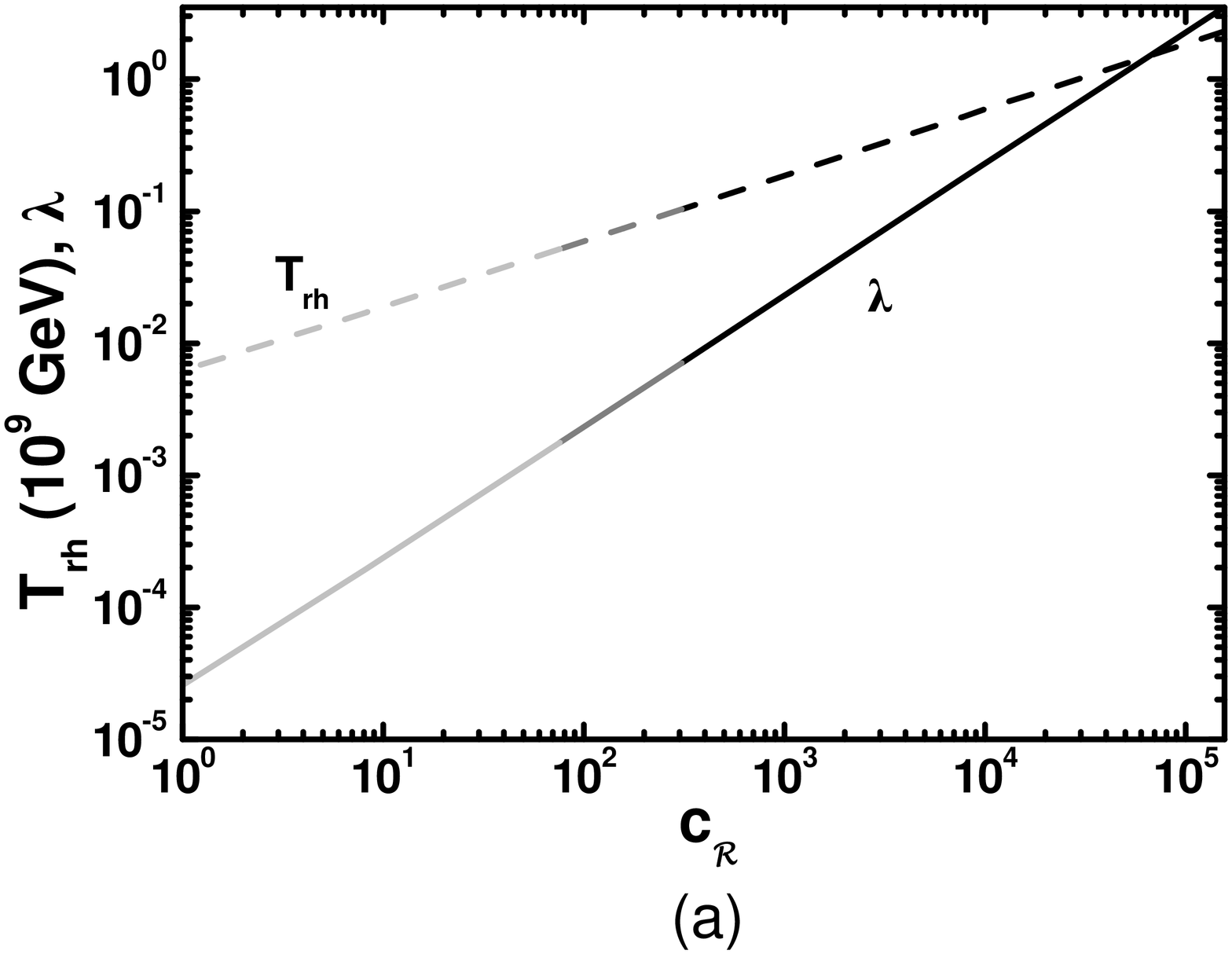,height=3.6in,angle=-90}
\hspace*{-1.2cm}
\epsfig{file=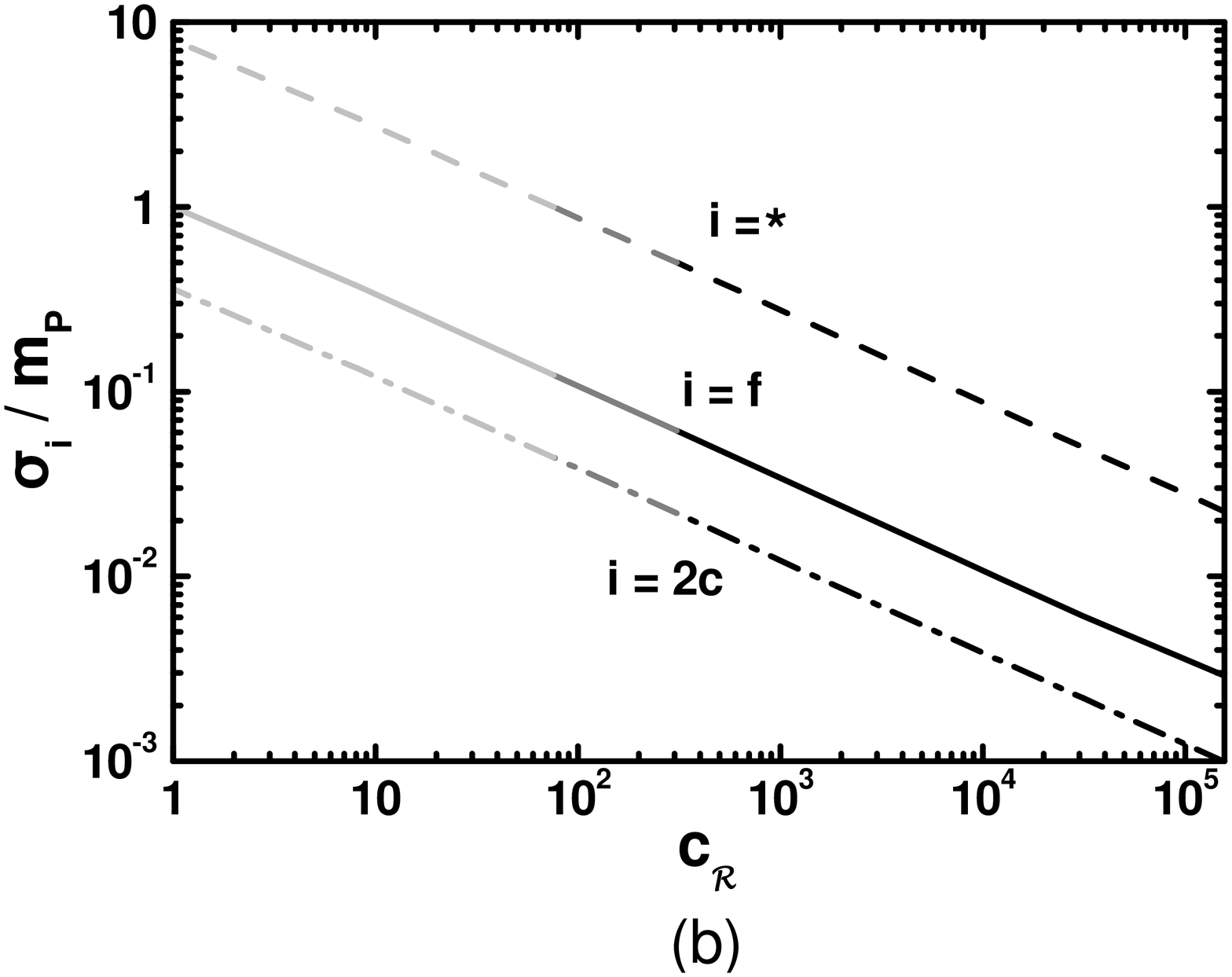,height=3.6in,angle=-90} \hfill
\end{minipage}
\hfill \vchcaption[]{\sl\small  The allowed by Eqs.~(\ref{Ntot}),
(\ref{obs}{\sf\ftn a}), (\ref{Vl}{\sf\ftn b}) and
(\ref{Vl}{\sf\ftn c}) values of $\ld$ (solid line) and $\Trh$ --
given by \Eref{T1rh} -- (dashed line) [$\sg_{\rm f}$ (solid line)
and $\sg_*$ (dashed line)] versus $\ck$ (a) [(b)] for
$\la=10^{-6},~\kx=1$ and $\hef=10^{-5}$. Also, $\sg_{2\rm
c}$ (dot-dashed line) given by \sEref{scr}{b} as function of
$\ck$ is shown. The light gray and gray segments denote values of the
various quantities satisfying \sEref{Vl}{a} too, whereas along the
light gray segments we obtain $\sg_*\geq\mP$.}\label{fig2}
\end{figure}

%%%%%%%%%%%%%%%%%%%%%%%%%%%%%%%%%%%%%%%%%%

Our results are presented in \Fref{fig2}, where we draw the
allowed values of $\ld$ (solid line) and $\Trh$ (dashed line)
[$\sg_{\rm f}$ (solid line) and $\sg_*$ (dashed line)] versus
$\ck$ (a) [(b)] for $\hef=10^{-5}$ and $f_a=10^{12}~\GeV$. In
\sFref{fig2}{b} we also draw $\sg_{\rm 2c}$ -- derived from
\sEref{scr}{b} -- as a function of $\ck$. The upper [lower] bound
on $\ck$ comes from the saturation of the inequality in
\sEref{Vl}{b} [\sEref{Vl}{c}]. On the other hand, \sEref{Vl}{a} is
valid along the gray and light gray segments of the various
curves. Along the light gray segments, though, we obtain
$\sg_*\geq\mP$. The latter regions of parameter space, although
can be considered as less favored, are not necessarily excluded
\cite{circ}, since the energy density of the inflaton remains
sub-Planckian and so, corrections from quantum gravity can be
assumed to be small. In all, we obtain
\beq\label{res1} 1\lesssim
\ck\lesssim1.56\cdot10^5\>\>\>\mbox{and}\>\>\>2.6\cdot10^{-5}\lesssim
\ld\lesssim3.5\>\>\>\mbox{for}\>\>\> 52.5\lesssim
\Ne_*\lesssim54.6.\eeq
From \sFref{fig2}{a}, we observe that $\ld$ depends on $\ck$
almost linearly. This can be understood by combining \Eref{Prob}
and \sEref{obs}{a}. The resulting relations reveal that $\ld$ is
to be proportional to $\ck$, so as \sEref{obs}{a} is satisfied
with almost constant $\Ne_*$. Indeed we find
\beq\ld={3\cdot10^{-4}\pi\ck/\Ne_*}\>\Rightarrow\>
\ck=41850\ld\>\>\>\mbox{for}\>\>\>\Ne_*\simeq55.\label{la}\eeq
On the other hand, the variation of $\sg_{\rm f}$ and $\sg_*$ as a
function of $\ck$ -- drawn in \sFref{fig2}{b} -- is consistent
with \eqs{s*}{sgap}. If $\ck$ varies within its allowed region as
given in \Eref{res1}, we obtain
\beq\label{res} 0.963\lesssim \ns\lesssim0.965,\>\>\>-6.8\lesssim
{\as\over10^{-4}}\lesssim-6.1\>\>\>\mbox{and}\>\>\>4.4\gtrsim
{r\over10^{-3}}\gtrsim3.4.\eeq
Clearly, the predicted $\ns$ and $r$ lie within the allowed ranges
given in \sEref{obs}{b} and \sEref{obs}{c} respectively, whereas $\as$
remains quite small. These findings depend very weakly on $\Trh$
-- and therefore on $\hef$, since this controls the value of $\Trh$
via \Eref{T1rh} --  because $\Trh$ appears in Eq.~(\ref{Ntot})
through the one third of its logarithm, and consequently its
variation upon some orders of magnitude has a minor impact on the
required value of $\Ne_*$.

%%%%%%%%%%%%%%%%%%%%%%%%%%%%%%%%%%%%%%%%%%%%%%%%%%%%%%%%%%%%%%%%%%%%
\begin{figure}[!t]\vspace*{-.46in}\begin{tabular}[!h]{cc}\begin{minipage}[t]{7in}
%\begin{center}
\hspace{0.5in}
\epsfig{file=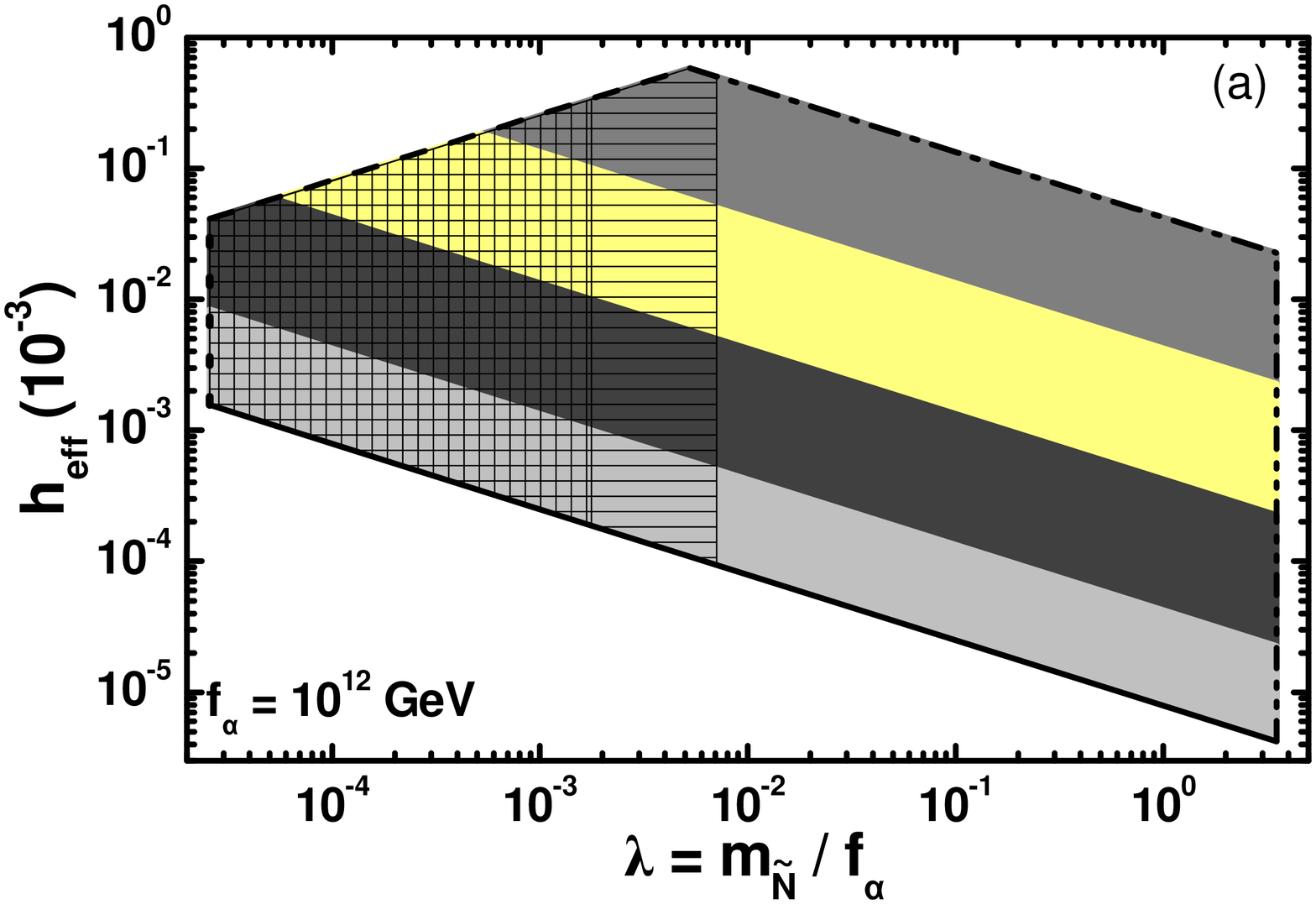,height=3.65in,angle=-90}\\
\hspace*{0.55in}\epsfig{file=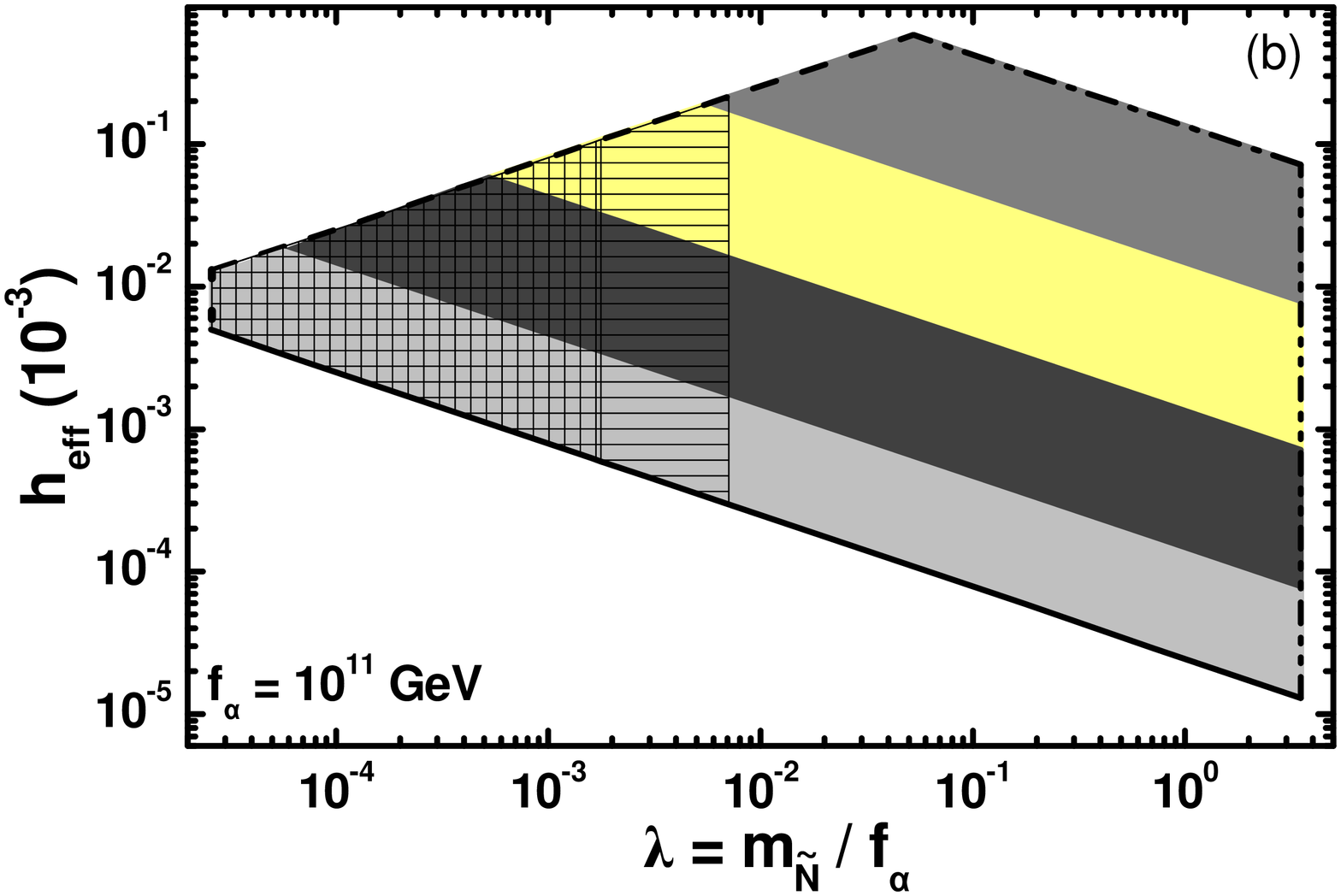,height=3.65in,angle=-90}\end{minipage}
&\begin{minipage}[h]{3in}
\hspace{-3.in}{\vspace*{-5.in}\includegraphics[height=5.65cm]
{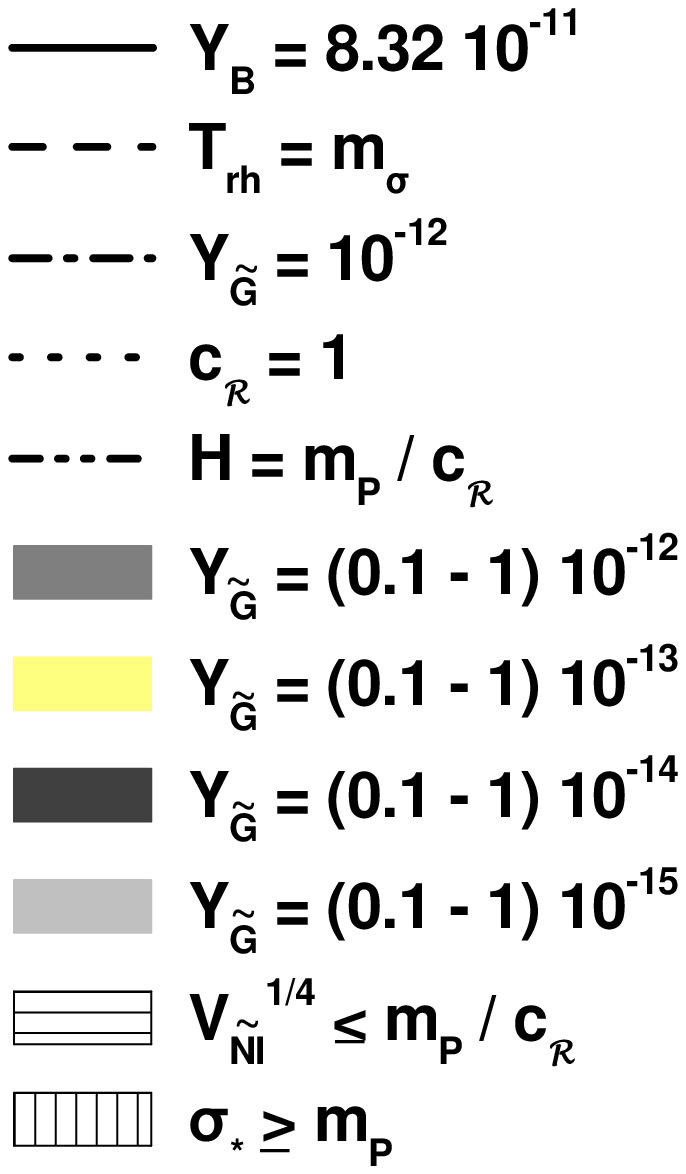}}\end{minipage}
\end{tabular} \hfill \vchcaption[]{\sl \small Allowed (shaded) regions
as determined by Eqs. (\ref{wash}), (\ref{Vl}{b}), (\ref{Vl}{c}),
(\ref{BAUwmap}) and (\ref{Ygw}) in the $\ld-\hef$ plane, for
$\la=10^{-6},~\kx=1,~\mu=1~\TeV$ and $f_a=10^{12}~\GeV$ (a) or
$f_a=10^{11}~\GeV$ (b). The conventions adopted for the various
lines and shaded or hatched regions are also shown.}\label{fig3}
\end{figure}
%%%%%%%%%%%%%%%%%%%%%%%%%%%%%%%%%%%%%%%%%

On the contrary, $\hef$  plays a key-role in simultaneously
satisfying \eqss{wash}{BAUwmap}{Ygw} -- see \eqs{Yb}{Ygr}. For
this reason we display in \sFref{fig3}{a} [\sFref{fig3}{b}] the
allowed regions by all imposed constraints in the $\ld-\hef$ plane
for $f_a=10^{12}~\GeV$ [$f_a=10^{11}~\GeV$] -- cf.
\cref{sneutrinoF}. The restrictions on the parameters arising from
the inflationary epoch are denoted by dotted and double-dotted
dashed lines, whereas the ones originating from the
post-inflationary era are depicted by solid, dashed and dot-dashed
lines. In particular, the double-dotted dashed [dotted] lines come
from the bounds of \sEref{Vl}{b} [\sEref{Vl}{c}]. In the
horizontally lined regions \sEref{Vl}{a} holds, whereas in the
vertically hatched region we get $\sg_*\geq\mP$. On the other
hand, the solid [dashed] lines correspond to the lower [most
conservative upper] bound on $Y_B$ [$\Yg$] in \Eref{BAUwmap}
[\Eref{Ygw}]. Since we use $|\deff|=1$, it is clear from
\eqs{el}{Yb} that values of $\hef$ above the solid line are
compatible with the current data in \Eref{BAUwmap} for
conveniently adjusing $|\deff|<1$. However the strength of $\hef$
can be restricted by the bounds of \Eref{Ygw}, which can be
translated into bounds on $\Trh$ via \Eref{Ygr}. Specifically we
obtain $Y_{\Gr}\simeq\lf0.1-1\rg\cdot10^{-12}$
or~$\Trh\simeq\lf0.53-5.3\rg\cdot10^9~\GeV$ (gray area),
$Y_{\Gr}\simeq\lf0.1-1\rg\cdot10^{-13}$
or~$\Trh\simeq\lf0.53-5.3\rg\cdot10^8~\GeV$ (yellow area),
$Y_{\Gr}\simeq\lf0.1-1\rg\cdot10^{-14}$
or~$\Trh\simeq\lf0.53-5.3\rg\cdot10^7~\GeV$ (dark gray area) and
$Y_{\Gr}\simeq\lf0.1-1\rg\cdot10^{-15}$
or~$\Trh\simeq\lf0.53-5.3\rg\cdot10^6~\GeV$ (light gray area). The
competition of the two restrictions above can be presented also
analytically. Indeed, plugging $\Trh$ via \Eref{T1rh} into
\eqs{Yb}{Ygr}, seting $Y_L=Y_L^{\rm min}$ and
$Y_{\Gr}=Y_{\Gr}^{\rm max}$ -- the exact numerical values can be
extraced by \eqs{BAUwmap}{Ygw} respectively -- and solving the
resulting equations w.r.t. $\hef$, we obtain the following
inequilities:
\beq \mbox{{\sf\ftn
(a)}}\>\>\>\hef\geq\frac{2\sqrt{\pi}}{\sqrt{\mP\msn}}{Y_L^{\rm
min}\over |c_L|c_T}\>\>\>\mbox{and}\>\>\>\mbox{{\sf\ftn
(b)}}\>\>\>\hef\leq\frac{2\sqrt{\pi}}{\sqrt{\mP\msn}}{Y_{\Gr}^{\rm
max}\over c_{\Gr}c_T}\cdot \label{hefb}\eeq
These reproduce quit accuratelly the behavior seen in
\Fref{fig3}. On the other hand, the out-of-equilibrum condition
depicted by a dashed line -- see \Eref{wash} -- puts the upper
bound on $\hef$ in a minor portion of the parameter space.

Comparing \sFref{fig3}{a} and \sFref{fig3}{b}, we conclude that
$\msn$ given by \sEref{masses}{a} is reduced for
$f_a=10^{11}~\GeV$ w.r.t. its value for $f_a=10^{12}~\GeV$, and so
the condition of \Eref{wash} cuts a larger slide of the available
parameter space. Letting $\ld$ vary within its allowed region in
\Eref{res1}, we obtain
\beq\label{res2}
4.3\cdot10^{-6}\>[1.3\cdot10^{-5}]\lesssim\hef/10^{-3}\lesssim0.58\>\>\>
\mbox{for}\>\>\>f_a=10^{12}~\GeV\>[f_a=10^{11}~\GeV]\eeq
%\\ \lesssim\hef/10^{-3}\lesssim0.58\>\>\>\mbox{and}\>\>\>\lesssim\deff\leq1
%\>\>\>\mbox{for}\>\>\eea
%
where the overall minimal [maximal] $\hef$ can be found in the
upper, almost central [lower right] corner of the allowed region.
As we see above and can be induced by Eqs.~(\ref{wash}) and
(\ref{hefb}{\sf\ftn b}), the maximum allowed $\hef$ is $f_a$
independent and it is obtained for
$\ld\simeq0.005~[\ld\simeq0.05]$ and
$f_a=10^{12}~\GeV~[f_a=10^{11}~\GeV]$. This point gives also a
lower bound on $|\deff|$,  $|\deff|\gtrsim2\cdot10^{-4}$ -- which
is obviously also $f_a$ independent. Note finally that in both
cases the resulting $\msn$'s can be much lower than those obtained
within the simplest model of sneutrino inflation with a quadratic
potential \cite{murayama, sneutrino1}. On the other hand, $\hef$
turns out to be comparable with the one obtained in those models.

\section{Conclusions}\label{con}

In this paper we attempted to embed within a realistic
cosmological setting one of the recently formulated \cite{linde2}
SUSY models of chaotic inflation with non-minimal coupling to
gravity. We concentrated on a moderate extension of the MSSM
augmented by three RH neutrino superfields and three other singlet
superfields, which lead to a PQPT tied to renormalizable
superpotential terms. The coupling between the RH neutrinos and
one of the fields associated with the PQPT plays a crucial role
for the implementation of our scenario. We showed that the model
non only supports non-MI driven by the lightest RH sneutrino, but
it also resolves the strong CP and the $\mu$ problems of the MSSM
and, even more, it leads to the production of the required by the
observations BAU via non-thermal leptogenesis, which accompanies
the inflaton's decay. Moreover the $\Gr$ abundance becomes
observationally safe for $\Gr$ masses even lower than $10~{\rm
TeV}$. An important prerequisite for all these is that the
parameter of the superpotential related to the PQPT, $\la$, is
adequately small. Imposing a number of observational constraints
arising from the data on the inflationary observables, the BAU,
the concentration of the unstable $\Gr$ at the onset of
nucleosynthesis and the mass of the heaviest light neutrino, we
restrict the effective Yukawa coupling, involved in the decay of
the inflaton, to relatively small values, and the inflaton mass to
values lower than $10^{12}~\GeV$.

\begin{acknowledgement} \paragraph{}

We would like to cordially thank G. Lazarides for helpful
discussions and J.~McDonald for an enlightening correspondence.

\end{acknowledgement}

\appendix
\setcounter{equation}{0}
\renewcommand{\theequation}{A.\arabic{equation}}
\renewcommand{\thesubsection}{A.\arabic{subsection}}
%\section*{Appendix}\section{Inflation with non-Minimal Gravitational Coupling in SUGRA}
%\section*{Appendix A: Non-Minimal Inflation in SUGRA}
\section*{Appendix: Non-Minimally Curvature-Coupled Scalars in SUGRA} \label{sugra}

Non-MI can be realized by a scalar field with a non-minimal coupling to
the Ricci scalar curvature. The formulation of a such theory
within SUGRA is described below. Recall that we follow the
conventions of \cref{kolb} for the quantities related to the
gravitational sector of our set-up.

\rhead[\fancyplain{}{ \bf \thepage}]{\fancyplain{}{\sl
Non-M$\sn$I, PQPT \& non-Thermal Leptogenesis}}
\lhead[\fancyplain{}{\sl
\hspace*{-.3cm}\leftmark}]{\fancyplain{}{\bf \thepage}} \cfoot{}

%\subsection{}\label{sugra1}

In contrast to the non-SUSY case -- see e.g. \cref{wmap3, sm1,
nmchaotic, nmi} -- we find it convenient to start our analysis
with the general \emph{Einstein-frame} (EF) action for the scalar
fields $\phi^\al$ plus gravity in four dimensional, ${\cal N}=1$
SUGRA \cite{linde1,linde2}:
\beq \label{action1} S=\int d^4x \sqrt{-\widehat
g}\lf-\frac{1}{2}\mP^2 \rce +K_{\al\bbet}\geu^{\mu\nu}
D_\mu\phi^\al D_\nu\phi^{*\bbet}-\Ve\rg, \eeq
where hat is used to denote quantities defined in the EF;
$\widehat g$ is the determinant of the Friedmann-Robertson-Walker
background metric \cite{kolb};
\beq K_{\al\bbet}=\frac{\partial^2
K}{\partial\phi^\al\partial\phi^{*\bbet}}>0\>\>\>\mbox{and}\>\>\>D_\mu\phi^\al=\partial_\mu\phi^\al-A^A_\mu
k^\al_A\eeq
are the covariant derivatives for scalar fields $\phi^\al$. Here
$A^A_\mu$ stand for the vector gauge fields and $k^\al_A$ is the Killing
vector, defining the gauge transformations of the scalars
\cite{linde2}. Assuming that the $D$-terms of $\phi^\al$ vanish --
as for the singlet scalars $\sn, P, X$ and $\bar X$ in our model
-- the EF scalar potential, $\Ve$, is given in terms of the
K\"ahler potential, $K$, and the superpotential, $W$, by
\begin{equation}
\Ve=e^{K/\mP^2}\left(K^{\al\bbet}F_\al F_\bbet-3\frac{\vert
W\vert^2}{\mP^2}\right),\label{Vsugra}
\end{equation}
with $K^{\al\bbet}K_{\al\beta}=\delta^\bbet_\beta$ and
$F_\al=W_{,\phi^\al} +K_{,\phi^\al}W/\mP^2$.

The action in \Eref{action1} can be brought the \emph{Jordan
frame} (JF) by performing a conformal transformation
\cite{conformal}. Indeed, if we define the JF metric,
$g_{\mu\nu}$, through the relation
\beq \label{weyl}
\geu_{\mu\nu}=-\frac{\Omega}{3}g_{\mu\nu}~~\Rightarrow~~\left\{\bem
%\begin{array}{rl}
\sqrt{-\geu}=\Omega^2\sqrt{-g}/9\>\>\>\mbox{and}\>\>\>
\geu^{\mu\nu}=-3g^{\mu\nu}/\Omega, \hfill \cr
\rce=-3\left(\rcc-\Box\ln \Omega+3g^{\mu\nu} \partial_\mu
\Omega\partial_\nu \Omega/2\Omega^2\right)/\Omega \hfill \cr\eem
%\end{array}
\right.\eeq
-- where $\Box=\lf
-g\rg^{-1/2}\partial_\mu\lf\sqrt{-g}\partial^\mu\rg$ -- we obtain
the action in the JF as follows
\beq S=\int d^4x \sqrt{-g}\lf\frac{\mP^2}{6}\Omega
\rcc+\frac{\mP^2}{4\Omega}\partial_\mu\Omega\partial^\mu\Omega
-\frac{1}{3}\Omega K_{\al{\bbet}}D_\mu\phi^\al D^\mu
\phi^{*\bbet}-V  \rg\>\>\>\mbox{with}\>\>\>V
=\frac{\Omega^2}{9}\Ve.\label{action2}\eeq
Taking into account that $\partial_\mu\Omega=D_\mu\Omega$ -- since
$\Omega$ is gauge invariant -- and that the purely bosonic part,
${\cal A}_\mu$, of the on-shell value of the auxiliary field
$A_\mu$ is given by
\beq {\cal A}_\mu =-\frac{i}{2\Omega}\mP^2\lf
D_\mu\phi^\al\Omega_\al-D_\mu\phi^{*\aal}\Omega_\aal\rg
\label{Acal}\eeq
-- with $\Omega_\al=\Omega_{,\phi^\al}$ and
$\Omega_\aal=\Omega_{,\phi^{*\aal}}$ -- and for the choice
\beq\Omega = -
3e^{-K/3\mP^2}\>\Rightarrow\>K=-3\mP^2\ln\lf-\Omega/3\rg,\label{Omg1}\eeq
we arrive at the following action
\beq S=\int d^4x \sqrt{-g}\lf\frac{\mP^2}{6}\Omega
\rcc+\mP^2\Omega_{\al{\bbet}}D_\mu\phi^\al
D^\mu\phi^{*\bbet}-\Omega {\cal A}_\mu{\cal A}^\mu/\mP^2-V
\label{Sfinal} \rg. \eeq
It is clear from the first term of the RHS of this expression
that the resulting $S$ exhibits non-minimal couplings of the
$\phi^\al$'s to $\rcc$. However, $\Omega$ enters in the kinetic
terms of the $\phi^\al$'s too. In order to get canonical kinetic terms,
we need $\Omega_{\al{\bbet}}=\delta_{\al{\bbet}}$ and ${\cal
A}_\mu=0$. The first condition is satisfied \cite{linde2} by the
choice
\beq \Omega= - 3 + \delta_{\al{\bbet}} {\phi^\al
{\phi}^{*\bbet}\over\mP^2} -3\big( {\rm F}(\phi^\al)+{\rm
F}^*(\phi^{*\aal})\big), \label{Omg}\eeq
where F is a dimensionless, holomorphic function, which
expresses the non-minimal coupling to gravity. Note
that even when ${\rm F}(\phi^\al)=0$ for some $\al$, the $\phi^\al$'s
are conformally coupled to gravity due to the second term of the
RHS of the expression above. This choice for the
frame function leads via \Eref{Omg1} to the
following K\"ahler potential
\beq K= - 3\mP^2 \ln \lf 1-\frac{1}{3\mP^2}\delta_{\al{\bbet}}
\phi^\al {\phi}^{*\bbet} +\lf {\rm F}(\phi^\al)+{\rm
F}^*(\phi^{*\aal})\rg\rg. \label{Kg}\eeq
On the other hand, ${\cal A}_\mu=0$ when the dynamics of
the $\phi^\al$'s is dominated only by the real moduli $|\phi^\al|$.
Therefore, it is possible to get a SUGRA realization of the
inflationary models with non-minimal coupling.

\rhead[\fancyplain{}{ \bf \thepage}]{\fancyplain{}{\sl
Non-M$\sn$I, PQPT \& non-Thermal Leptogenesis}}
\lhead[\fancyplain{}{\sl \leftmark}]{\fancyplain{}{\bf \thepage}}
\cfoot{}

\end{document}